\documentclass[]{Photos_interface_design}
\usepackage{graphicx}
\usepackage{hyperref}
\usepackage{eurosym}
\usepackage{pifont}
\usepackage{amsmath}
\usepackage{subfigure}
\usepackage{booktabs}

\makeatletter
\renewcommand*\l@subsection{\@dottedtocline{2}{1.5em}{2.0em}}
\renewcommand*\l@subsubsection{\@dottedtocline{3}{3.0em}{3.0em}}
\makeatother

\begin{document}

\maketitle

\tableofcontents\pdfbookmark[0]{Table of Contents}{toc}

\newpage

\section{Introduction}
For a long time, {\tt PHOTOS} Monte Carlo \cite{Barberio:1990ms,Barberio:1993qi} 
has been used for the generation of bremsstrahlung in the decay of particles and resonances.
Over the years the program has acquired
popularity and it evolved into a high 
precision tool \cite{Golonka:2006tw}. Since 2005, when
multi-photon radiation was 
introduced \cite{Golonka:2005pn} into the program (version 2.15), there were no 
further public upgrades 
of the program until 2010. The efforts were concentrated on documentation and 
new tests;
 phase space treatment was shown to be 
exact \cite{Nanava:2006vv} and for several 
processes \cite{Golonka:2006tw,Nanava:2006vv,Nanava:2009vg}
an exact matrix element was studied with the help of optional weights.
Benchmark distributions, including comparisons with  
other simulation programs, were collected on the {\tt MC-TESTER}~\cite{Davidson:2008ma} (special program devoted to tests) web page \cite{Photos_tests}%
\footnote{An up-to-date version of the {\tt PHOTOS} code described in this paper is
available from the web page of our project~\cite{photosC++}.}. 

 Such high precision applications require good control of the event record content on which {\tt PHOTOS} operates. On one side it 
requests skills and experience of the user and on the other it provides 
the flexibility necessary for the study of effects like, for example, systematic errors for 
measurements of anomalous couplings or $W$ cross section. Methods of 
correlated samples  can be applied\footnote{To exploit such methods in 
the high precision regime, good control of matrix element properties is necessary.
As was shown in \cite{Kleiss:1990jv}, complications for such methods arise at the second order matrix element only, thus at the precision level of 
$(\frac{\alpha_{QED}}{\pi})^2 \simeq 10^{-5}$.}. 

Until 2010 the {\tt HEPEVT} \cite{Altarelli:1989wu} event record was used as the structure for 
communication between physics Monte Carlo programs and detector/reconstruction 
packages. Experimental physicists used {\tt HEPEVT} 
for their own applications  as well. Later, to gain  flexibility, {\tt FORTRAN} was replaced by {\tt C++} and 
instead of {\tt HEPEVT}, the {\tt C++} event structure  {\tt HepMC} \cite{Dobbs:2001ck}
was used. Nothing prevented 
moving {\tt PHOTOS} to a {\tt C++} environment, allowing the use of event records such as {\tt HepMC},
and to rewrite the whole 
of {\tt PHOTOS} to {\tt C++}. In fact implementation of the algorithm in that language 
is clearer and easier to
 maintain. Because of its design the {\tt PHOTOS} algorithm benefits from the object 
oriented features of {\tt C++}. It is our third program, after {\tt MC-TESTER} \cite{Davidson:2008ma}
and the {\tt TAUOLA} interface \cite{Davidson:2010rw}, already previously ported to {\tt HepMC} and {\tt C++}.
This completes the main step of migration of these three programs to the new style.

Such migration is convenient for the users too; they can now work
with  homogeneous {\tt C++} software. From the physics point of view, transformation 
of {\tt PHOTOS} 
from {\tt FORTRAN} to {\tt C++}  brings some benefits as well.
The channel dependent, complete first order matrix elements of {\tt PHOTOS}, in {\tt FORTRAN},
 are available only 
in special
kinematical configurations. With the help of the new  event record interface they  become
available for general use.
For that purpose, better access to the information necessary to orient the spin state of decaying particles
is now provided.

Our paper is organized as follows. Section \ref{sec:requrements} is devoted
to a description of the physics information which must be available in the event
record for the algorithm to function. Later, particular requirements for the 
information stored in {\tt HepMC} are given. Section \ref{sec:design} describes
the program structure. In Section \ref{sec:extensibility} methods prepared for 
extensions to improve the physics precision of the generator are explained.
Section \ref{sec:tests} presents the program tests and benchmarks. 
A summary, section \ref{sec:summary}, closes the paper.
There are three appendices \ref{Interface to PHOTOS}, 
\ref{sec:User Guide} and \ref{sec:User Configuration} attached to the paper.
Respectively, they describe the interface to the old part of the code, which has been rewritten from {\tt FORTRAN} to {\tt C},
provide a user guide and explain the program configuration methods and parameters. 

This document concentrates on features of {\tt PHOTOS} version 3.60, for 
other versions, consult {\tt README} files and {\tt changelog.txt} of the {\tt documentation/doxygen} directory.

\section{Requirements of the {\tt PHOTOS} Interface}
\label{sec:requrements}
The algorithm of {\tt PHOTOS} Monte Carlo can be divided into two parts.
The first, internal part, operates on elementary decays. Thanks to carefully 
studied properties of the 
QED (scalar QED), the  algorithm (with certain probability) 
replaces the kinematical configuration of the Born level decay with a new one, 
where a bremsstrahlung photon, photons or lepton pairs
are added and other particle momenta are modified. This part of the program is sophisticated from the physics 
point of view \cite{Golonka:2006tw,Nanava:2006vv},
but from the point of view of data structures the algorithm is simple.
That is why the gain from re-writing this part of the program to {\tt C} is rather
limited. Nonetheless, there were no obstacles for such a transformation to be
performed and it is completed now. In fact it was already done
previously \cite{photosplus}, but the resulting program was developed too early 
and did not attract users because of a lack of a {\tt C++} event record format standard at that time.

The typical result of high energy process simulation are events of complex structure.
They include, for example, initial state parton showers, hard scattering parts,
hadronization and finally chains of cascade decays of resonances. 
A structure similar to a tree is created, but properties of such data structures
are sometimes violated.
For its action, {\tt PHOTOS} needs to scan an event record (tree) 
and localize decays (branches in the tree) where
it is supposed to act. The decaying particle (mother) and its primary decay products
(daughters) have to be passed into the internal event structure of {\tt PHOTOS}.
For calculation of matrix element kernels, mothers of the decaying particle are needed as well. 
Finally for each daughter a list of all its subsequent decay products has to be 
formed. Kinematical modifications need to be performed on all descendants of the modified daughter.

In the new {\tt C++} version of the event record part of the algorithm, additional functionality
is added.
The first mother of the decaying particle will be localized and passed together with  
the elementary branching%
\footnote{We will use branching to refer to decay of particle or resonance (usually, but not always, represented by a decay vertex) which {\tt PHOTOS} can process. }
 to the internal part of the program. 
Prior to activation of the algorithm for  photon(s) (and/or lepton pairs) generation and kinematic construction,
 the whole decay branching 
(supplemented with its mother(s))
will be boosted into the decaying particle's rest frame and the first mother
will be oriented along the $z$ axis. 
In many cases, the spin state of the decaying particle  can be calculated from kinematics of its production process.
Later it is passed on, to the code which calculates the matrix element for the branching.

The part of the code responsible for photon(s) (lepton pair) 
generation and kinematic 
construction has been also rewritten from {\tt FORTRAN} to {\tt C}. It has been extended
over the last five years and features the options presented above.

Before an actual description of the program, let us list the tasks the event record interface must be able to perform:
\begin{enumerate}
\item a method to read  particles stored in the event tree.
\item a method to add or modify particles of the event tree.
\item a method to search for elementary decays over the entire tree of the event.
\item a method to form lists of all subsequent decay products originating from each elementary decay product.
\item a method to localize the first mother of the decaying particle. 
\item a method to localize the second mother for the special case of a $t \bar t$ pair.
\end{enumerate}

\subsection{ Requirements specific to event records}

The {\tt C++} version of the {\tt PHOTOS} interface implements all functionality
of its predecessor, {\tt PHOTOS} version 2.15 \cite{Golonka:2005pn} coded in {\tt FORTRAN}.
The program has the process-dependent correcting weights of refs 
\cite{Golonka:2006tw,Nanava:2009vg} installed.
{\tt PHOTOS} can be attached to any Monte-Carlo program,
provided that its output is available through an event record for which an interface has been provided.
The default distribution of {\tt PHOTOS} contains an interface for the {\tt HepMC} \cite{Dobbs:2001ck} event record,
as well as an interface for an old {\tt FORTRAN} event record, {\tt HEPEVT}%
\footnote{This standard is less commonly used, thus the interface to it is less tested.}.

It seems that {\tt HepMC} will
remain a generally accepted standard for the near future. However,
already now several different options for how {\tt HepMC} is used are
widespread. The possibility of the flexible  adaptation of our event record 
interface to different
options has been considered in the design,  drawing experience
from {\tt MC-TESTER} \cite{Davidson:2008ma,Golonka:2002rz}.

\subsection{Object Oriented Event Records  -- The Case of {\tt HepMC}}
 In adapting the {\tt PHOTOS} interface to the {\tt C++} event record format
the difference between the {\tt HEPEVT} event record, with its variant still used
by the core of the {\tt PHOTOS} code (as a struct type),
and the {\tt HepMC} event record has to be taken into account.  In the first case 
a {\tt FORTRAN common block} containing a list of particles with their properties and
with integer variables denoting pointers to their origins and
descendants is used.  The {\tt HepMC} event structure is built from vertices,
each of them having pointers to their origins and descendants. Links
between vertices represent particles (or fields).  Fortunately, in both {\tt
  FORTRAN} and {\tt C++} cases, the event is structured as a
tree\footnote{At least in principle, because in practice its
properties may be rather like a graph without universally defined
properties.  This makes our task challenging.}; the necessary
algorithms are analogous, but nonetheless different. The {\tt HepMC}
structure based on vertices is more convenient for the {\tt PHOTOS}
interface. 

In {\tt HepMC}, an event is represented by a {\tt GenEvent} object,
which contains information such as event id,
units used for dimensional quantities in the event and the list of produced particles. The particles
themselves are grouped into {\tt GenVertex} objects allowing access to mother
and daughter particles of a single decay. Vertices provide an easy way
to point to the whole branch in a decay tree that needs to be accessed,
modified or deleted if needed. The information of a particle  itself is stored
in a {\tt GenParticle} object containing the particle id, status and momentum
as well as information needed to locate its position in the decay tree.
This approach allows traversing the event record structure in several different
ways.

The {\tt HepMC} event record format is  evolving with time, making it necessary
 to adapt
the code to new versions. The
{\tt HepMC} versions 2.06, 2.05  and 2.03 were used  in the final tests of our 
distribution%
\footnote{The interface has also been tested and is fully compatible
with the alpha4 version of {\tt HepMC} 3.0}.

Evolution of the {\tt HepMC} format itself is not a crucial problem.
In contrast, conventions on how physics information is  filled into {\tt HepMC}
represent the main source of technical and also physics 
challenges for our interface. 
This is quite similar to the previous
{\tt HEPEVT - FORTRAN} case. Let us discuss this point in more detail now.

\subsubsection{Event Record Structure Scenarios}

While many Monte-Carlo generators (e.g. {\tt PYTHIA 8.1} \cite{Sjostrand:2007gs}, 
{\tt HERWIG++} \cite{Bahr:2008pv}), {\tt SHERPA} \cite{Gleisberg:2008ta} can 
store events in {\tt HepMC} format, the  representations of
these events are not subject to strict standards,  and can therefore
vary between Monte-Carlo generators or even physics processes. Some examples
of these variations include the conventions of status codes, the  way
documentary information on the event is added, the direction of pointers at a vertex
and the conservation (or lack of conservation) of energy-momentum at a vertex.
Below is a list of properties for basic scenario we have observed in Monte-Carlo
generators used for testing the code.

This list will serve as a declaration for the conventions of  {\tt HepMC} filling, which  the 
interface should interpret correctly.

\begin{itemize}
  \item \textbf{4-momentum conservation} is assumed for all vertices in the event record where {\tt PHOTOS} is expected to act.
  \item \textbf{Status codes:} only information on whether a given particle is decaying (status 2) or stable (status 1) is used.
  \item \textbf{Pointers at a vertex} are assumed to be bi-directional. 
    That is, the record structure may be traversed from mother to daughter 
    and from daughter to mother along the same path.
\end{itemize}

\noindent
\textbf{ Extensions/Exceptions} to these specifications are handled in some cases. We will call them
options for conventions of event record filling.
  \begin{itemize} 
    \item  Vertices like $\tau^\pm \rightarrow \tau^\pm$ and $\tau^\mp \rightarrow \tau^\mp \gamma$ 
           where the decaying particle flavor is among its decay products will prevent  {\tt PHOTOS} being invoked.

    \item  If there is  4-momentum non-conservation\footnote{For details see 
           Appendix~\ref{subsection:other_methods}.} in the vertex,
           {\tt PHOTOS} will not be invoked too.  A special kinematic correcting
           method to remove smaller inconsistencies resulting e.g. from 
           rounding errors is available, but it must be used carefully to avoid
           action on vertices where four-momentum is not conserved because 
           of physics reasons.

    \item
           As in the {\tt FORTRAN} cases, we expect that  new  types of 
           conventions for filling the event record
           will appear, because of physics motivated requirements.
           Unfortunately, the resulting options do not always guarantee
           an algebraically closed structure.  
           Host program-specific patches may need to be defined for
           {\tt PHOTOS}. 
           Debugging can then be time consuming, and will need to be repeated for every new
           case.
           
    \item  In the case of low-mass particles that are vulnerable to numerical fluctuation (such as muons and electrons),
           the correct information about a particle's mass is expected. Since in such cases the mass calculated
           from a 4-vector can often be incorrect (including negative values). An appropriate method
           of {\tt PHOTOS} must be used to correct that information (see \ref{subsection:other_methods}).
   \end{itemize}

 In the future,  an important special case of event record's filling, with
information extracted from experimentally observed events (e.g. $Z\to \mu^+\mu^-$
 modified later to $Z\to \tau^+\tau^-$) should be allowed.
  Obviously, a new type (or types) of {\tt HepMC} filling will then appear.
The possibility of reverse operation of {\tt PHOTOS}, actually to remove 
photons from the decay vertex
may be then envisaged.

A good example for event record divergence from the standard,
is the evolution of {\tt PYTHIA}. While in version 8.108 the status codes for 
our example processes took the values 0, 1 or 2  only (in the part of the record 
important for {\tt PHOTOS}), other values were already present in
version 8.135. As a consequence the status code for 
otherwise nicely decaying particles was not always 2 anymore. We have introduced 
a change  in the file PhotosEvent.cxx to adjust. After  the change
the program should work for all previous cases as before, 
changes like this one are usually difficult to validate
and complicated  tests are necessary. One could  investigate 
if instead of changes to the {\tt PHOTOS} algorithm a different choice of  input for {\tt PYTHIA} would not 
be a more appropriate 
solution, but in this case we choose to adapt our algorithm%
\footnote{ At present, our programs, {\tt TAUOLA} and
 {\tt PHOTOS}, supplement the event record with new particle entries carrying bar codes 
with values starting from 10001. That is the choice resulting from our use 
of {\tt HepMC} methods and defaults.  }.

\subsection{Interface to the Event Record stuct of {\tt HEPEVT} type}
\label{sect:F77fill}

It was rather simple to  rewrite {\tt PHOTOS} to
{\tt C++} completely. To profit from numerical tests, the  core of {\tt PHOTOS}
was rewritten to essentially  plain  {\tt C}. Common blocks were replaced
with structs, old names of methods and functions were preserved. The 
{\tt C++} part of the code searches the whole event for
suitable {\tt HepMC} vertices for the generation of bremsstrahlung. Once such
a vertex is found it is copied to an internal event record struct  which is 
called {\tt hep}  (in older versions it was {\tt FORTRAN} {\tt PH$\_$HEPEVT} common block);
it uses {\tt HEPEVT}  as a specification for struct type definition.
The {\tt C} code of {\tt PHOTOS} is then executed.
The data structure passed in this way is rather simple. Only a single vertex consisting
of the decaying particle along with its mothers and daughters is passed. Information 
on mothers is nececcary  for the calculation of process dependent, matrix element based, 
kernels.

A description of the interface to the internal, essentially plain {\tt C}, parts of the code is
given in  Appendix \ref{Interface to PHOTOS}.

\section{Design}
\label{sec:design}
\subsection{Classes and Responsibilities}

The choice of splitting the source code into three main modules
 allows the separation of the {\tt C}  code of the numerical algorithm from the abstract {\tt C++} interface
and the concrete implementation of the interface created for the appropriate
event record.

\begin{figure}[h!]
\centering
\includegraphics[scale=0.5]{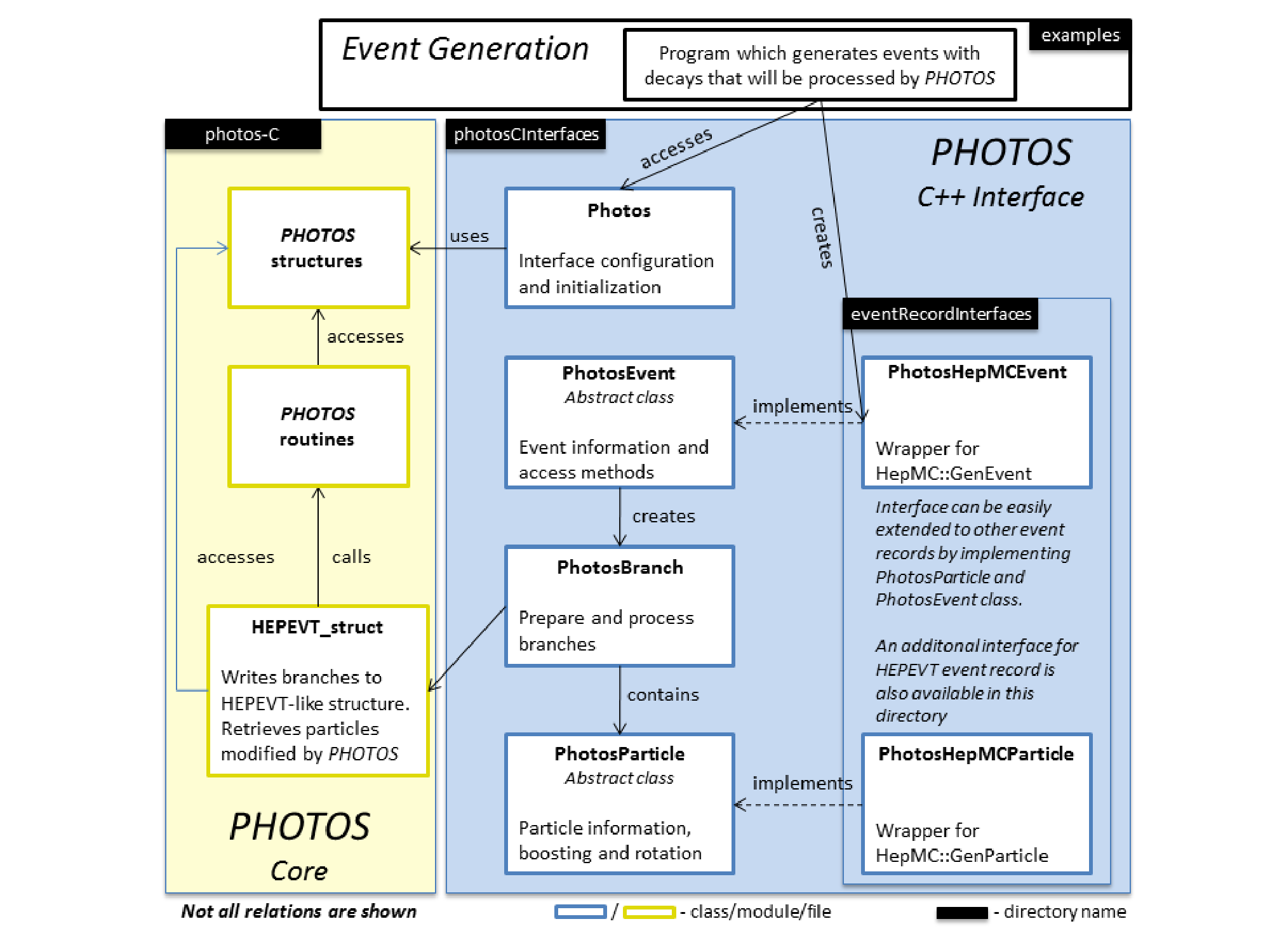}
\label{fig:design}
\caption{{\tt PHOTOS} {\tt C++} interface class relation diagram}
\end{figure}

\begin{itemize}
  \item {\bf {\tt photos-C} }\\
       This part of the code includes numerical algorithms of {\tt PHOTOS}. In particular,
       the code for generating photons as well as electron-positron and muon pairs.
       The {\tt HEPEVT\_struct} class located here is responsible for writing
       the decay branch to be processed into the internal event record struct {\tt hep} as
       well as for writing the output of the processing back to the event record.
       For further details see Appendix \ref{Interface to PHOTOS}.
  \item {\bf {\tt PHOTOS} {\tt C++} Interface} \\
       This is an abstract interface to the event record.
       The class {\tt PhotosEvent} contains information regarding the whole event
       structure, while {\tt PhotosParticle} stores all information regarding a single particle.
       All particles used by the interface are located in the event in the form of
       a list of {\tt PhotosParticle} objects.
       The last class located here, {\tt PhotosBranch}, contains information regarding
       elementary branching to be processed by {\tt PHOTOS}. In particular, 
       it contains the algorithm
       which selects single branching for processing and filters out branchings that will not be processed.
  \item {\bf Event Record Interface} \\
       This contains the event record implementation classes. All classes stored here represent
       the implementation of specific event record interfaces and are responsible for reading,
       traversing and writing to the event record structure.
       Only the {\tt PhotosEvent} and {\tt PhotosParticle} classes must be implemented.
       The {\tt HepMC} event record interface is implemented
       through {\tt PhotosHepMCEvent} and {\tt PhotosHepMCParticle}. These classes are similar to the
       analogous {\tt TAUOLA} \cite{Davidson:2010rw} event record classes.
       An example of a minimalistic interface to an event record has been provided
       through the classes {\tt PhotosHEPEVTEvent} and {\tt PhotosHEPEVTParticle}\footnote{This interface is the 
       only way of implementing NLO corrections in programs using the {\tt HEPEVT} event record.
       Let us note that for this part of the code only minimal set of  tests was completed, and its use should be restricted, until further tests.}.
       For example, a separate set of classes deriving from {\tt PhotosParticle} and {\tt PhotosEvent}
       can be written for {\tt HepMC} events generated by {\tt PYTHIA}, treating particle status codes
       differently than before, and a separate set for {\tt HepMC} events taken from other MC generators.
       The choice between these two implementations can be introduced by the user's code.
\end{itemize}

\subsection{Directory Structure}

\begin{itemize}
\item {\bf src/eventRecordInterfaces/ } - source code for classes which interface with event records.
      Currently, the {\tt HepMC} interface and an interface to the {\tt FORTRAN} event record {\tt HEPEVT} are located here.\\
  Classes:
  \begin{itemize}
  \item { \bf PhotosHepMCEvent} - interface to HepMC::GenEvent objects. 
  \item { \bf PhotosHepMCParticle} - interface to HepMC::GenParticle objects. 
  \item { \bf PhotosHEPEVTEvent} - interface to the event structure of the {\tt HEPEVT} format, used in the interface to the {\tt HEPEVT} common block of {\tt FORTRAN} example only. 
  \item { \bf PhotosHEPEVTParticle} - interface to a single particle from the {\tt HEPEVT} event record, used in the interface to the {\tt HEPEVT} common block of {\tt FORTRAN} example only.
  \end{itemize}   

\item {\bf src/photosCInterfaces/ } - source code for the abstract event record interface.  \\
  Classes:
  \begin{itemize}
  \item { \bf Photos } - controls the configuration and initialization of {\tt PHOTOS}.
  \item { \bf PhotosEvent } - abstract base class for event information.
  \item { \bf PhotosParticle } - abstract base class for particles in the event.
  \item { \bf PhotosBranch } - contains one {\tt PhotosParticle} and  pointers to its mothers and daughters.
                               The main algorithms of the abstract interface, such as invoking {\tt PHOTOS} processing
                               or filtering out branchings that will not be processed, is defined here.
  \end{itemize}

\item {\bf src/utilities/ } - source code for utilities.\\
  Files/classes:
  \begin{itemize}
  \item { \bf Log} - general purpose logging class that allows filtering out output messages 
      of the {\tt PHOTOS {\tt C++} Interface} and tracks statistics for each run.
  \item { \bf PhotosRandom} - random number generator from Ref.~\cite{James:1988vf,marsaglia:1987} taken from {\tt PHOTOS FORTRAN} and rewritten to {\tt C++}.
  \item { \bf PhotosDebugRandom} - static class derived from {\tt PhotosRandom} that provides several tools to store and restore
                                   the state of the {\tt PHOTOS} random number generator.
  \item { \bf PhotosUtilities.cxx} - support functions (e.g. boosting, rotations, etc.)
  \end{itemize}

\item {\bf  src/photos-C/ } - core {\tt PHOTOS} code. Since version 3.54, {\tt PHOTOS} has been
      fully rewritten to {\tt C++} and located in its own namespace {\tt Photospp}\footnote{This means that no part of
      the code is shared with old {\tt PHOTOS FORTRAN} and both versions can be loaded simultanously without the
      risk of one version overwriting the options of the other.}. For algorithmic backward
 compatibility\footnote{The resulting modules are however not interchangeable and the program
 will not function if the {\tt PHOTOS FORTRAN} library is loaded instead of the code encapsulated 
in the {\bf src/photos-C/ } directory.
}
 the code structure as in {\tt FORTRAN} version is kept. The appropriate descriptions remain valid;
in publications as well as in the code, now in {\tt C++}. \\
  Files:
  \begin{itemize}
    \item { \bf photos-C.cxx } - core functionality of {\tt PHOTOS}. Structures of this code contain internal variables such as weights, angles, four momenta, etc.
                                 The {\tt Photos} class uses some of these structures to set several initialization options.
    \item { \bf HEPEVT\_struct.cxx } - static class translating information about particles passed through the abstract event record interface
                                       to a {\tt HEPEVT}-like structure used by the core {\tt PHOTOS} code.
    \item { \bf forW-MEc.cxx } - routines to calculate the matrix element in $W$ decays. 
                                 Note distinct programming style from the rest of code in the {\tt photos-C} directory.
    \item { \bf forZ-MEc.cxx } - routines to calculate the matrix element in $Z$ decays. 
                                 Note distinct programming style from the rest of code in the {\tt photos-C} directory.
    \item { \bf pairs.cxx } - code for electron-positron and muon pairs emission.
    \end{itemize}

  \item {\bf examples/ } - examples of different {\tt PHOTOS} {\tt C++} Interface uses.
    \begin{itemize}
        \item {\bf photos\_hepevt\_example} - stand alone example with a simple 
        $e^+e^- \rightarrow \tau^+\tau^-$ event written in {\tt HEPEVT} format
         and then processed by {\tt PHOTOS}.
	\item {\bf photos\_standalone\_example} - the most basic example which loads pre-generated 
	      events stored in a file in {\tt HepMC} format which are then processed by {\tt PHOTOS}.
	\item {\bf single\_photos\_gun\_example} - an example of using the {\tt processOne} method
	      to process only selected vertices within the event record.
    \item {\bf photos\_pythia\_example} - an example of $e^+e^- \rightarrow Z \rightarrow \mu^+\mu^-$ events
	generated by {\tt PYTHIA 8} and processed by {\tt PHOTOS}. The analysis is done using {\tt MC-TESTER}, initialization for emission of lepton pairs
is demonstrated.
   \item {\bf photosLCG\_pythia\_example} - similar to previous case, prepared
to demonstrate LCG scripts.
    \item {\bf tauola\_photos\_pythia\_example } - an example of  {\tt TAUOLA} linked with {\tt PYTHIA 8}.
	The decay chain is processed by {\tt PHOTOS} and then analyzed with {\tt MC-TESTER}.
\item {\bf testing/photos\_tauola\_test} - test program, may be useful as an example for the user's own work, but rather not as an introductory example. See {\tt README} files of the directory. 
\item {\bf testing/photos\_test} - test program, may be useful as an example for user own work, but rather not as an introductory example. See {\tt README} files of the directory. An example of pair emission use is given.
\item {\bf testing/further\_subdirectories} directories with numerical benchmark results, see Subsection \ref{section:BenchmarkFiles}.
    \end{itemize}   
  \item {\bf include/} - directory for the header files.
  \item {\bf lib/ } - directory for the compiled  libraries. 
  \item {\bf documentation/ } - contains doxygen documentation and this latex file.
\end{itemize}

\subsection{Algorithm Outline}
\label{sect:Outline}

An overview of the algorithm for  the {\tt PHOTOS C++ Interface} is
given below. For clarity, the example assumes that the processed event
is stored in the {\tt HepMC} event record structure.

The first step is creation of a {\tt PhotosHepMCEvent} object from
a {\tt HepMC::GenEvent} event record. After that, the {\tt process()} method should
be executed by the user's code\footnote{Instead of creating a {\tt PhotosHepMCEvent} and processing the whole event,
a user may want to execute {\tt Photos::processParticle(...)} or {\tt Photos::processBranch(...)}
on the single branching  or branch where {\tt PHOTOS} is expected to perform its tasks.
For details see Appendix~\ref{PHOTOSgun}.
}, invoking the following process:

\begin{enumerate}
\item The {\tt HepMC} event record is traversed and a list of all decaying
      particles is created.
\item Each particle is checked and if the resulting branching is a self-decay\footnote{A history entry in the event record, like
      $Z\to Z$ or $\tau \to \tau$.} it is skipped.
\item For each remaining particle a branch,  including the particle's mothers and daughters
      is created. Special cases consisting of mothers and daughters but without  intermediate particle 
to be decayed are also added to the 
	  list of branches.
\item Branchings are filtered out again, this time with  the user's choice of processes
      to be skipped by {\tt PHOTOS}.
\item Each branching is processed by {\tt PHOTOS} separately:

	\begin{enumerate}
  
	\item The branch is written to a {\tt hep} struct.
	\item The {\tt PHOTOS} photon and pair adding algorithm is invoked. 
	\item The resulting branching is taken back from {\tt hep} and introduced to the event record.
	\item Any changes made by {\tt PHOTOS} to the already existing particles
          invokes kinematical changes to their whole decay trees.
	\item Finally, the created particles are added to the event record.
	\end{enumerate}

\end{enumerate}

The underlying {\tt HepMC::GenEvent} is hence modified with the insertion of new particles.
Alternatively, as in our example, {\tt PHOTOS/examples/photos\_hepevt\_example.f}, the interface {\tt HEPEVT} in {\tt FORTRAN} is used 
and then the content of  {\tt HEPEVT} is modified.

\section{Extensibility}
\label{sec:extensibility}
 The first purpose of the {\tt C++} interface to the {\tt C} {\tt PHOTOS} algorithm is to make available 
all the functionality of its {\tt FORTRAN} predecessor for 
{\tt C++} data
structures. Some new methods for improved initialization are introduced. The new program 
functionality has been prepared to enable extensions, such as emission kernels based on matrix elements
or emission of pairs. 
Let us briefly discuss some of these points.

\subsection{{\tt PHOTOS} Extensions}
So far  we have presented an algorithm, as it was 
already implemented in {\tt FORTRAN}. Further extensions were introduced:  

\begin{itemize}

\item
We have prepared the structure (branching including particle's mothers)  
for the implementation of channel dependent matrix elements. This was   
integrated into the {\tt C++} version of {\tt PHOTOS}. 

\item
Methods devised to check the content of the event record are described in Appendix \ref{App:Logging}. 
They need to be used whenever {\tt PHOTOS} 
processes events from a new generator e.g. upgraded versions of  {\tt PYTHIA},
which may fill the event record in an unexpected way.
Experience gained from many years of developing and maintaining the algorithm
have shown that this is the most demanding task; the necessity to
adapt to varying physics and technical inputs of the event record pose
a multitude of problems. The nature of these difficulties cannot be
predicted in advance. 

\item
For the sake of debugging we have introduced new control methods 
and ones which activate
internal printouts of the {\tt C} part of the code.
The routine {\tt PHLUPA} \cite{Barberio:1993qi} can be activated  to verify 
how an event is constructed/modified, and to investigate energy 
momentum (non-) conservation or other inconsistencies.
This is quite convenient, for example, for tracing problems in the
information passed to {\tt PHOTOS}.

\item
Numerical stability is another consideration; it cannot be separated from
physics constraints. The condition  $E^2-p^2=m^2$ may be broken  because of 
rounding errors.  However, due to intermediate particles with
  substantial widths, the on-mass-shell condition may not be applicable.
{\tt PHOTOS} may be adapted to such varying conditions, but it requires
good interaction with users. The protection which removes 
inconsistencies in the event record may be a source of unexpected difficulties
in other cases. 
\end{itemize}

\subsection{Event Record Interface}
In the times of {\tt FORTRAN}, the {\tt PHOTOS} interface used an internal event structure which was
based on {\tt HEPEVT},
adding to it (understood as a data type) an extra variable defining 
the status of particles with respect to QED Bremsstrahlung. We still use event objects based on {\tt HEPEVT} 
but they are declared as {\tt C} structs and used internally in the {\tt PHOTOS} algorithm.

In some cases, like
$\tau \to l \nu_l \nu_\tau$, bremsstrahlung was already generated earlier
by other generator, and {\tt PHOTOS} should not be active on such decays.
At present,  a set of initialization methods is 
prepared as described in Appendix \ref{section:suppress}. This superseeds 
the role of QED emission flags of {\tt FORTRAN} times. 

There is definite room for 
improvement. For example if the vertex $q \bar q \to l^\pm l^\mp g$ is encountered
(note the presence of $g$ in the final state),
the interface could `on the fly' add an intermediate $Z$ into the record and enable {\tt PHOTOS}
on the temporarily constructed decay branching, $Z \to l^\pm l^\mp $. 
We can process $q\bar q \to l\bar l$ without an intermediate $Z$ though.

Internally, in the {\tt C} part of {\tt PHOTOS}, the data
structs of {\tt C}  based on {\tt HEPEVT}: {\tt pho} and {\tt hep} 
are used, but they store only a single elementary decay. 
This solution  prevented
the need to redo many of the {\tt FORTRAN} era benchmarks.

\section{Testing}
\label{sec:tests} 
 Some of the most important parts of the {\tt PHOTOS} project are its physics oriented tests.
Several domains
of physics tests should be mentioned. Users interested in precision 
simulations of $Z$ or $W$ decays  will 
find  the papers \cite{Nanava:2009vg,Golonka:2006tw,Golonka:2005pn}
most interesting. There, careful comparison with the first order matrix element 
and confirmation of the agreement was shown.
For $Z$ decays, comparisons  with a Monte Carlo program based on exclusive 
exponentiation with up to the second order matrix element
is possible and was performed on some benchmark distributions.
Inclusion of a correcting weight for complete first order matrix elements was found to be numerically less important
than the absence of the second order matrix element in the YFS exponentiation scheme used by the reference programs. 
In these comparisons the Monte Carlo programs from LEP 
\cite{koralz4:1994,kkcpc:1999} were used as the reference. In numerical tests {\tt MC-TESTER} \cite{Davidson:2008ma}
was used. The advantage of the method is that {\tt C++} and {\tt FORTRAN} program
results can be easily compared\footnote{We thank Andy Buckley for checking numerically
 that our conclusions on the first order exact YFS exponentiation results extend
to the programs presently used at the LHC such as  
SHERPA and HERWIG++.  }.
 
For inclusive calculations, FSR radiative corrections are at the one permille level.
For semi-inclusive cross sections, such as the total rates of $Z$ decay for events
where the hardest photon energy (or two hardest photon energies)   exceed 1 GeV in the $Z$ rest frame, differences
between results from {\tt PHOTOS}, with the matrix element correcting weight turned off, and
and YFS based generators of the first or second order were also 
at the 0.2 \% level. 
On the other hand, if two  hard photons were requested and invariant masses constructed from leptons
and hard photons were monitored,
 the level of differences exceeded 
 30 \%. However, even in this region of phase space, {\tt PHOTOS},  without the correcting
weight, performs better\footnote{In 
  the phase space region where only one hard photon is tagged this conclusion seems to depend
  on the variant of exponentiation in use, \cite{koralz4:1994} or \cite{kkcpc:1999}.
                                }
than programs based on exponentiation and the first order matrix element only. 

This conclusion needs to be investigated if   
realistic experimental 
cuts are applied. Fortunately the necessary programs are available for $Z$ decay.
In the case of $W$ decay, second order Monte Carlo generators supplemented with 
exponentiation are not available at this moment.
 
For users interested in the simulation of
background for Higgs searches at the LHC and for any other applications where 
two hard photon configurations are important, studies based on the comparison with 
a double photon matrix element are of 
interest. For {\tt PHOTOS} Monte Carlo such tests were initiated 
in refs.~\cite{Barberio:1993qi,RichterWas:1994ep,RichterWas:1993ta}.
Finally, users interested in low energy processes where the underlying physics model 
for photon emission cannot be controlled by theory sufficiently well
(scalar QED may 
be considered only as the starting point), will profit 
from \cite{Nanava:2006vv,Nanava:2009vg}. In all cases it is important that
the  program generation cover the full phase-space and that there are no 
approximations in phase-space. As in the {\tt FORTRAN} version, the code 
features approximation in the kernel. In some cases the process dependent 
complete first order 
kernel is available. At present such an option is prepared 
(see Section \ref{sect:F77fill}) for $W$ and $Z$ decays. It is also available for the  decays of scalars 
into two scalars. Then, exact means exact with respect to scalar QED only. 

The main purpose of the present paper is program documentation. This is why
we also need to cover the program tests that guarantee its proper installation.
The physics tests discussed above 
do not guarantee that the program will perform well on a particular platform and installation. Tests and debugging of the installation
are necessary too.  If the content of the event record is non-standard or rounding errors are large, the performance of {\tt PHOTOS} will deteriorate.

The first check after installation of {\tt PHOTOS} is whether some energy momentum 
non-conservation appears. Such offending events should be studied
before {\tt PHOTOS}  and after {\tt PHOTOS} is run to modify them.
If it is impossible to understand why inconsistencies for energy momentum 
non-conservation were created by {\tt PHOTOS}, the authors should be contacted. Sometimes
monitoring how an event is constructed inside the internal {\tt C} part of the code
may be useful. For that purpose a monitoring option\footnote{See Appendix \ref{App:Logging}
for the command {\tt Log::LogPhlupa(int from, int to)}}
 is
available from the {\tt C++} part of the code. In practice only rather
 advanced users will profit from the printouts. However, it may be sent to
the authors and help to quickly identify the cause of the problems.

The next step in benchmarking relies on comparisons with reference distributions. 
At present, we store these tests using {\tt ROOT} \cite{Antcheva:2009zz} and our {\tt MC-TESTER} program \cite{Davidson:2008ma}.

\subsection{{\tt MC-TESTER} Benchmark Files}
\label{section:BenchmarkFiles}

Over years of development of the {\tt TAUOLA} and {\tt PHOTOS} programs a certain level 
of the automation of tests was achieved. It was found that monitoring all the invariant mass distributions which can be constructed out of a given decay represent 
a quite restrictive but easy to implement test.
Finding the relative fraction of events for each distinct final state 
 complemented that test and is implemented now in the public version of {\tt MC-TESTER}. 
We have applied this method 
for {\tt PHOTOS} too. In this case, some soft final state particles have to be ignored because we are bound to  work with  samples which otherwise would
exhibit properties of unphysical infrared regulators (see Section 6.1 of 
ref \cite{Davidson:2008ma} for more details). For the most popular 
decays the benchmarks are collected on our project web page \cite{Photos_tests}.
In our distribution, we have collected numerical results in the directory
{\tt examples/testing} and its subdirectories:  
{\tt Htautau, pairs, ScalNLO, ttbar, Wenu, Wmunu, WmunuNLO, Zee, Zmumu, 
ZmumuNLO} and {\tt Ztautau}. Each of them includes
an appropriate initialization file for the particular run of {\tt PYTHIA}. Numerical results from long runs of {\tt MC-TESTER} based tests
are stored for reference\footnote{Details on the initialization for the 
runs are given in 
{\tt README-benchfiles}.}. At present, our choice of tests is oriented toward 
the LHC user and radiative corrections in decay of $W$'s, $Z$'s and Higgs particles.
Most users at low energy experiments use the {\tt FORTRAN} version 
of the code, which is why our tests and examples for the {\tt C++} version are not geared toward 
low energy applications yet.

\subsection{Results}
\label{sec:results}
\begin{figure}[h!]
\centering
\subfigure[bremsstrahlung from  $\tau^+ $ decay only]{
\includegraphics[scale=0.35]{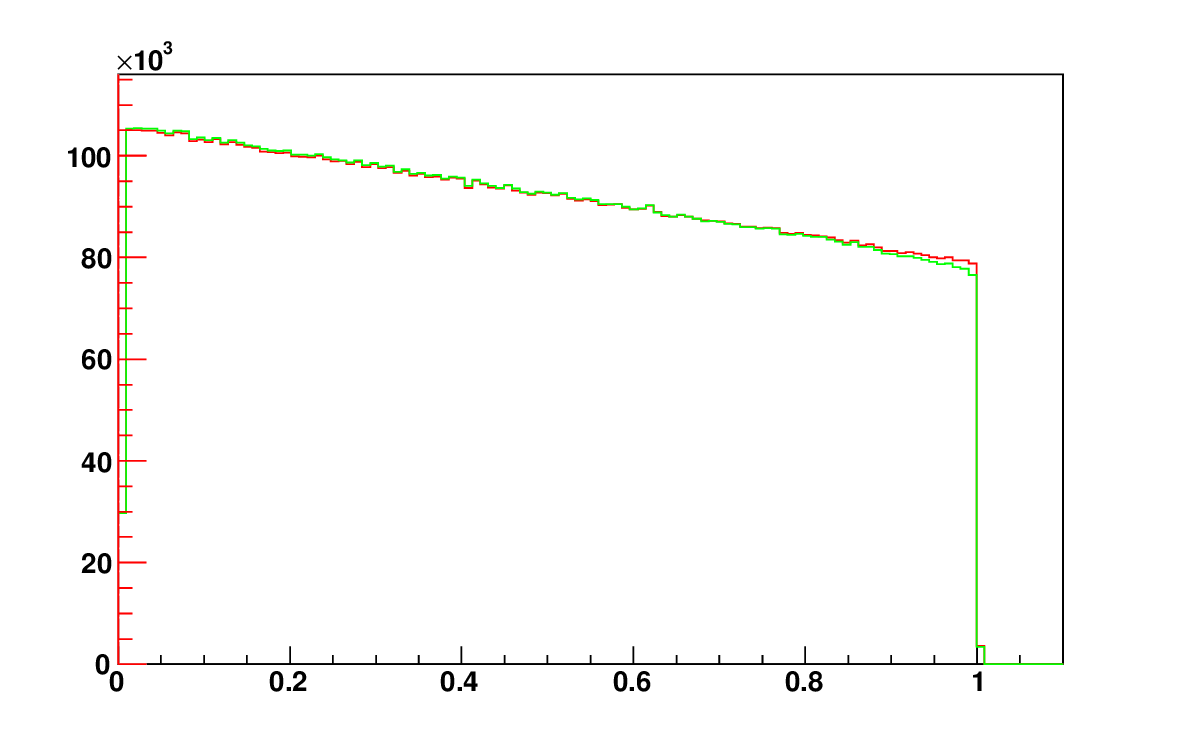}
}
\subfigure[bremsstrahlung from $Z$, $\tau^+ $ and $\tau^- $ decays]{
\includegraphics[scale=0.35]{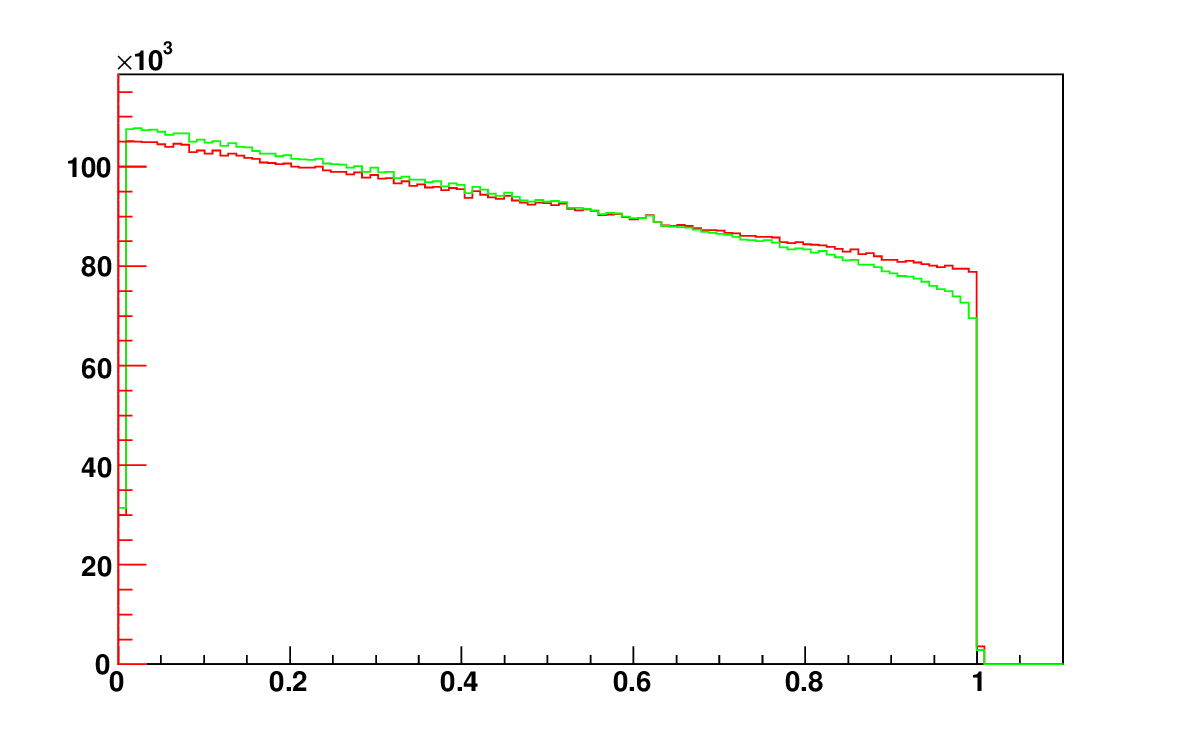}
}
\caption{ Bremsstrahlung effects for longitudinal spin observables
for the cascade decay: $Z \to \tau^+ \tau^-$, $\tau^\pm \to \pi^\pm\nu$.
The $\pi^+$ energy spectrum in the $Z$ rest-frame  is shown. The red line is for 
bremsstrahlung switched off
and green (light grey) when its effect is included. 
In the left plot, bremsstrahlung is in $\tau^+ $ decay only.
In the right plot, bremsstrahlung from $Z$ and  $\tau^\pm$ decays is
taken into account.
These plots have been prepared using a custom {\tt UserTreeAnalysis} of {\tt MC-TESTER}.
They  can be recreated with the test located in the {\tt examples/testing/Ztautau} directory, see  {\tt examples/testing/README-plots} for technical details. Results are 
consistent with Fig.~5 of Ref.~\cite{Eberhard:1989ve}.
\label{fig:KKMC}
}
\end{figure}

In principle, for the algorithm performing photon(s) construction, the {\tt C++} interface of
{\tt PHOTOS} introduces nothing new with respect to the 
version of {\tt PHOTOS} available in {\tt FORTRAN}.
That is why the  tests which are collected in \cite{Photos_tests} are not
recalled here, they were only reproduced and found to be consistent with the ones for
old {\tt PHOTOS FORTRAN}.
However the algorithm for combining a modified branching, including the added 
photon(s), to the rest of the event tree was rewritten.
Examples of spin dependent observables are of more interest because they test this new part of the code.
The $\pi$ energy spectrum in the decaying
Z boson rest-frame is the first example. 
The $\pi^\pm$ originates from a $\tau^\pm \to \pi^\pm \nu $ decay and, 
as was already shown a long time ago \cite{Boillot:1988re}, its energy spectrum is modified by bremsstrahlung both in $\tau$ and $Z$ decays. The net
bremsstrahlung  effect is similar to the one of e.g. $Z$ polarization. In Fig.~\ref{fig:KKMC} this result is reproduced.

\begin{figure}[h!]
\centering
\subfigure[selection C]{
\includegraphics[scale=0.35]{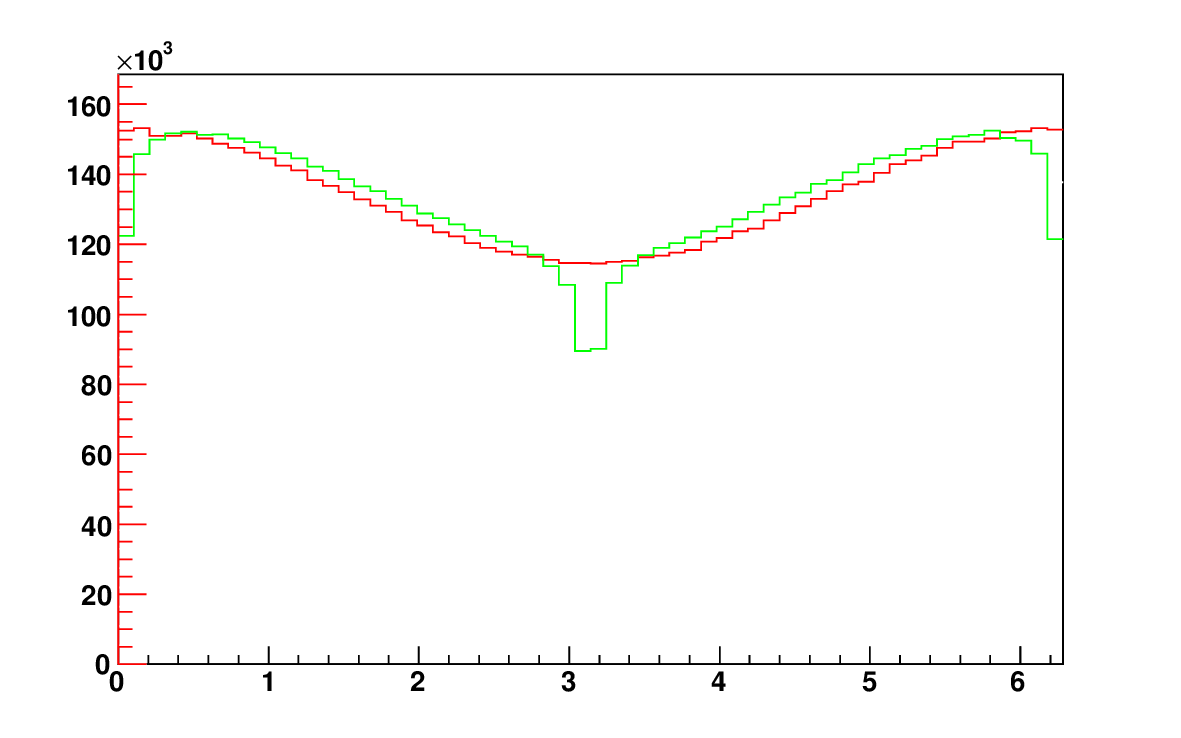}
}
\subfigure[selection D]{
\includegraphics[scale=0.35]{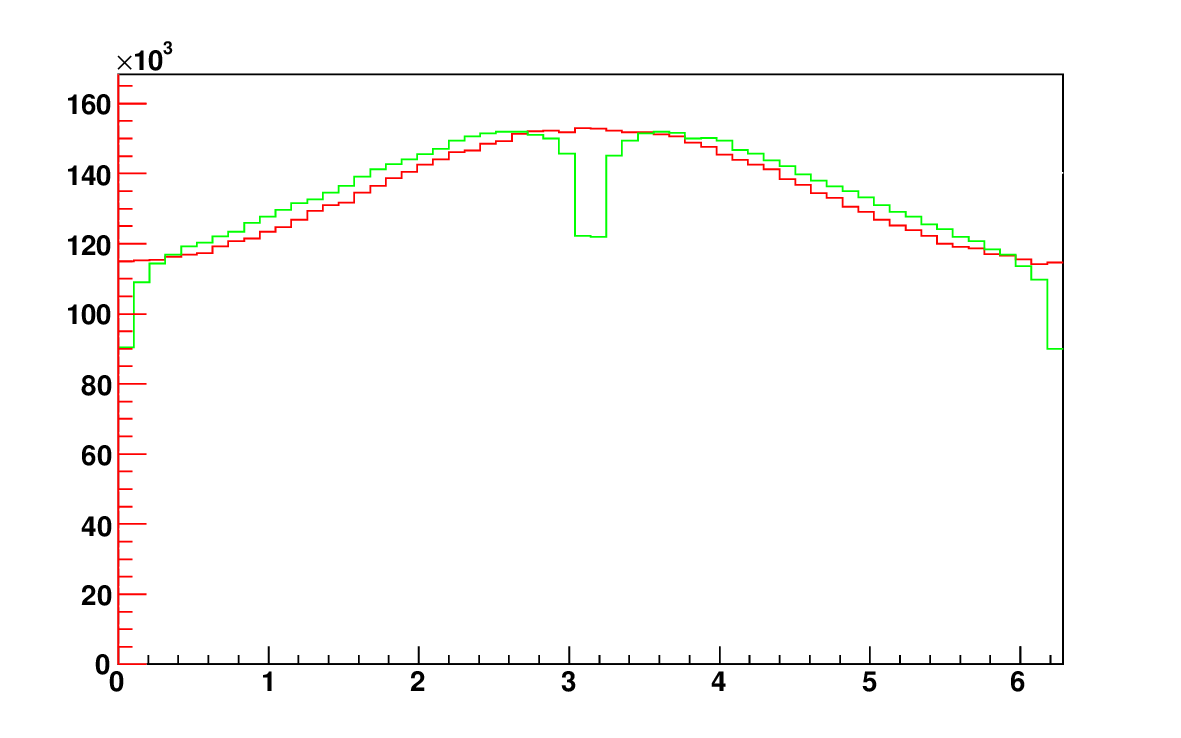}
}
\caption{Bremsstrahlung effects for transverse spin observable: 
  The distribution of the acoplanarity angle of oriented planes spanned respectively on 
  the $\pi^+\rho^+$ and $\pi^-\rho^-$ momenta is shown.  
The distribution is defined in the rest frame of the
  $\rho^+ \rho^-$ pair for the scalar Higgs decay chain $H\to
  \tau^+\tau^-$, $\tau^\pm \to \rho^\pm \nu_\tau$, $\rho^\pm \to
  \pi^\pm \pi^0$. {\tt PHOTOS} is used to generate
  bremsstrahlung.  The red curve indicates the distribution when
  bremsstrahlung effects are ignored and for the green curve (light grey) 
only events
  with bremsstrahlung  photons of energy larger than 1 GeV
  in the $H$ rest frame are taken. For the definition of selections C
  and D see.~\cite{Bower:2002zx,Desch:2003rw}.  These plots have been created using
  a custom {\tt UserTreeAnalysis} of {\tt MC-TESTER}.  They can be
  recreated by  the test located in the {\tt
    examples/testing/Htautau} directory, see {\tt
    examples/testing/README-plots} for technical details.
\label{fig:acoplanarity}
}
\end{figure}

Let us now turn to tests using observables which are sensitive to
transverse spin effects.  For this purpose we study the decay chain:
$H\to \tau^+\tau^-$, $\tau^\pm \to \rho^\pm \nu_\tau$, $\rho^\pm \to
\pi^\pm \pi^0$, where {\tt PHOTOS} may act on any of the branchings
listed above. An inappropriate action of the {\tt C++} part of {\tt PHOTOS}
could result in faulty kinematics, manifesting itself in a failure of
energy-momentum conservation or faulty spin
correlation sensitive distributions. However, as we can see from Fig.~\ref{fig:acoplanarity},
 the distributions (as they should be) remain nearly identical to the ones given in
\cite{Davidson:2010rw,tauolaC++}. The emission of soft and/or
collinear photons to the $\tau^+$ or $\tau^-$ does not change the
effects of spin correlations. The kinematical effects of hard,
non collinear  photons are responsible for dips in 
the acoplanarity distributions at $0$, $\pi$ and $2\pi$.


\begin{figure}[h!]
\centering
\includegraphics[scale=0.85]{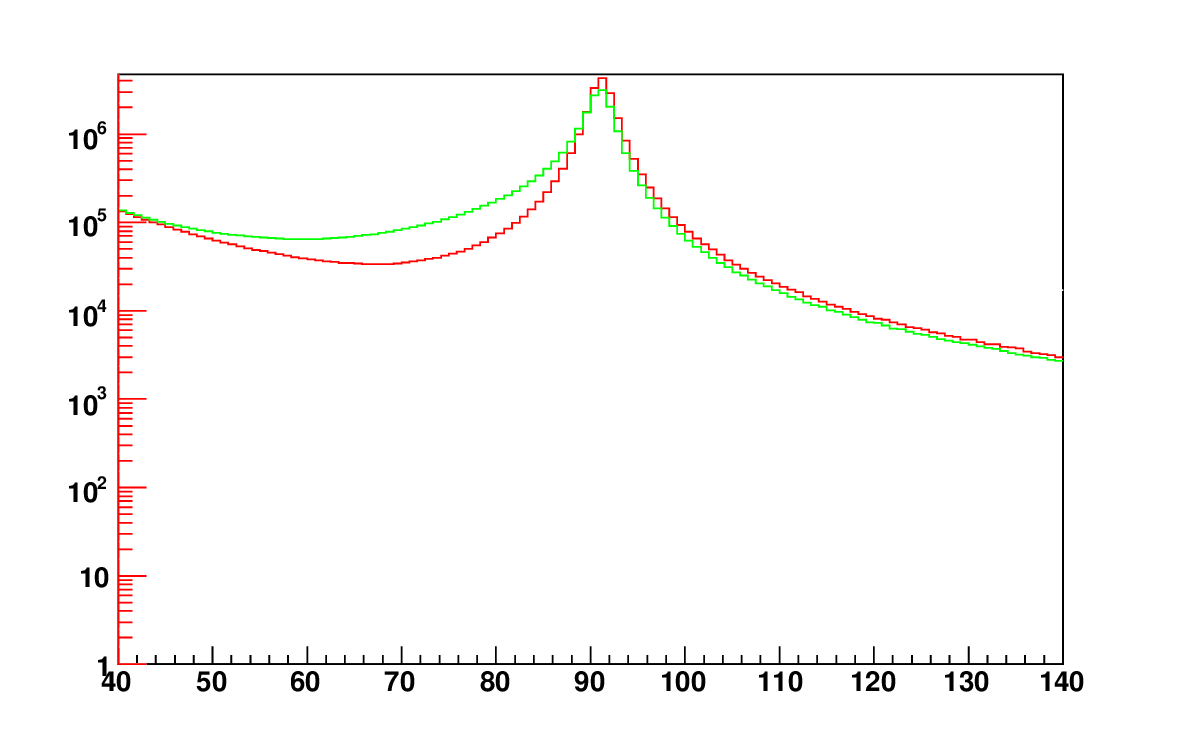}
\caption{Distribution of the bare $e^+e^-$ invariant mass. The green curve 
(light grey) represents results when final state 
bremsstrahlung is generated (with the help of {\tt PHOTOS} Monte Carlo). For the red curve FSR 
bremsstrahlung is absent.
The simulation of pp collisions at 14 TeV center-of-mass energy is performed using {\tt PYTHIA 8.1}.
The $Z/\gamma$ mediated hard process is used.
This plot has been created using a custom {\tt UserTreeAnalysis} of {\tt MC-TESTER}.
It can be recreated with the test located in the {\tt examples/testing/Zee} directory, see  {\tt examples/testing/README-plots} for technical details.
\label{fig:lineshape}
}
\end{figure}

In Ref.~\cite{Adam:2008ge} a discussion of the systematic errors for the measurement of the $Z$ cross 
section at the LHC is presented. One of the technical tests of our software is to obtain
Fig. 1b of that paper. In our Fig.~\ref{fig:lineshape} we have 
reproduced that plot using {\tt PYTHIA 8.1} and {\tt PHOTOS}. Qualitatively, the effect
of FSR QED bremsstrahlung is quite similar in the two cases.

In  Fig.~\ref{fig:lineshape} we present a plot of the bare electron pair 
(which means electrons are not combined with the collinear photons that accompany them) from $Z$
 decay with and without {\tt PHOTOS}. It is similar to
the plots shown for {\tt Horace} or {\tt Winhac}; see Refs.~\cite{CarloniCalame:2003ux,Winhac} and related
studies.
One should bear in mind that this is again a technical test with little 
direct application to physics. As explained in the figure caption, the LHC production process was used.

We have also checked that {\tt PHOTOS} works for $t \bar t$ events and the simulations explained in \cite{RichterWas:1993ta} can be repeated.
For this purpose we have produced $t \bar t$ pairs in $pp$ collisions at 
14 TeV center-of-mass energy. We have produced rates for events with zero, one or
at least two photons of energy above 0.001 of the $t \bar t$ pair mass
(energies are calculated in  the hard scattering frame).
Results are given in the following table which is constructed from  
$gg \to t \bar t$ events only:

\vspace{0.3cm} 
\begin{center}
{ \begin{tabular}{c c} 
\toprule 
Final state &  Branching Ratio (\%) $\pm$ Statistical Errors (\%) \\  
\midrule
{$ \widetilde{t} t \; \;\; \;$}  &  {99.0601 $\pm$ 0.0315}  \\ 
 {$  \widetilde{t} t \gamma \;\;$} &   { 0.9340 $\pm$  0.0031}   \\ 
{$  \widetilde{t} t \gamma \gamma$}  &  { 0.0060 $\pm$  0.0002}  \\ 
\bottomrule
\end{tabular} 
}  
\end{center} 

10 million events were generated and a slightly modified 
version of {\tt MC-TESTER}'s {\tt LC-analysis} from Ref.~\cite{Golonka:2002rz}
was used for calculation of the event rates\footnote{  We do not supplement the list of 
final state particles with the second mother, in contrast to the choice used in {\tt LC-analysis} from  Ref.~\cite{Golonka:2002rz}. }.
{\tt ROOT} files for differential distributions are 
collected in the directory {\tt examples/testing/ttbar}. 

\subsection{Tests Relevant for Physics Precision}

Let us now turn to an example of the test for the two photon final state configuration.
We compare (i) {\tt KKMC} \cite{kkcpc:1999} with exponentiation and second order matrix element ({\tt CEEX2}), (ii) {\tt KKMC} with exponentiation and first order matrix element ({\tt CEEX1})
and finally (iii) the results of {\tt PHOTOS} with exponentiation activated. As one can see from the table, the rates 
coincide for the three cases up to two permille for the event configurations 
where zero, one or at least two photons of energy above 1 GeV accompany the $\mu^+\mu^-$ pair.

\begin{table}
\centering 
\begin{tabular}{lrrr} 
\toprule 
Decay channel &\multicolumn{3}{c}{ Branching Ratio $\pm$ Statistical Errors   (100M event samples)} \\ 
      & {\tt KKMC CEEX2} (\%) & {\tt KKMC CEEX1} (\%)& {\tt PHOTOS} exp. (\%)\\ 
\midrule
 {$Z^{0} \rightarrow \mu^{+} \mu^{-} $} & {83.9190 $\pm$  0.0092} &{  83.7841 $\pm$  0.0092} & 83.8470 $\pm$ 0.0092\\ 
 {$Z^{0} \rightarrow \gamma \mu^{+} \mu^{-} $} & {14.8152 $\pm$  0.0038} &{  14.8792 $\pm$  0.0039} & 14.8589 $\pm$ 0.0039 \\ 
{$Z^{0} \rightarrow \gamma \gamma \mu^{+} \mu^{-} $} & { 1.2658 $\pm$  0.0011} &{   1.3367 $\pm$  0.0012} & 1.2940 $\pm$ 0.0011\\ 
\bottomrule
\end{tabular}
\end{table}

\begin{figure}[h!]
\centering
\subfigure[{\tt CEEX2}: red; {\tt CEEX1}: green]{
\includegraphics[scale=0.35]{{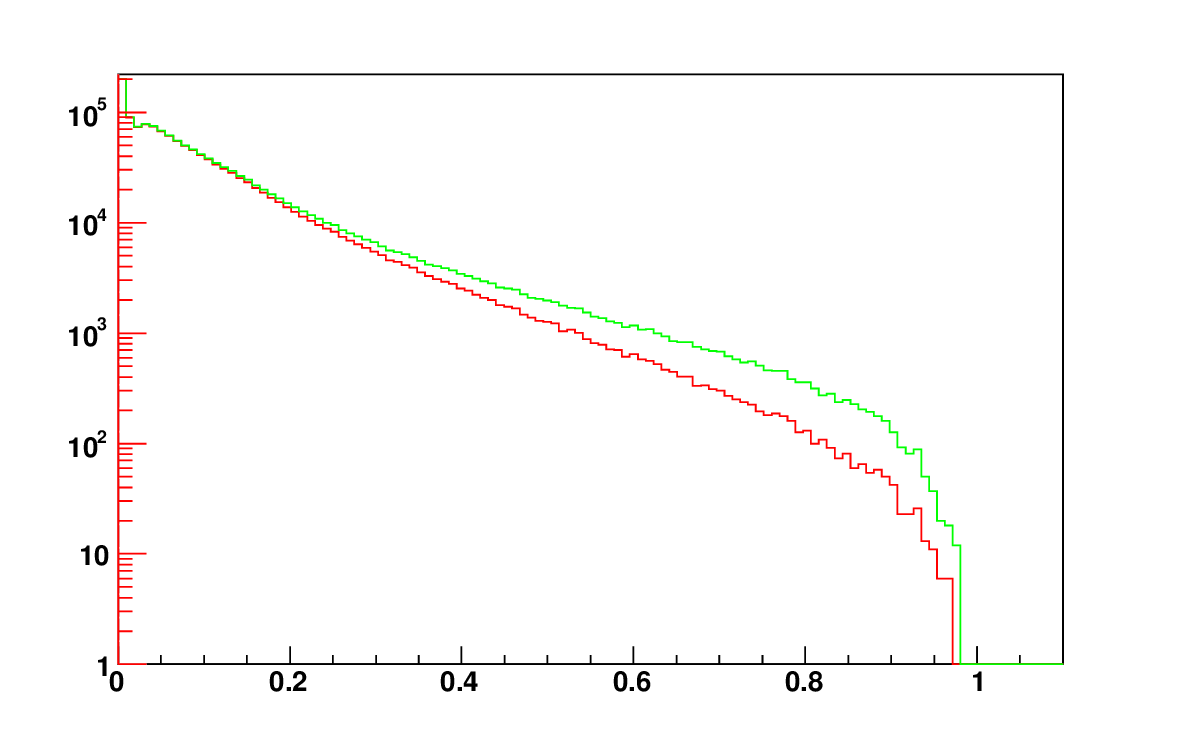}}
}
\subfigure[ {\tt CEEX2}: red; {\tt PHOTOS}: green]{\label{plot:b}
\includegraphics[scale=0.35]{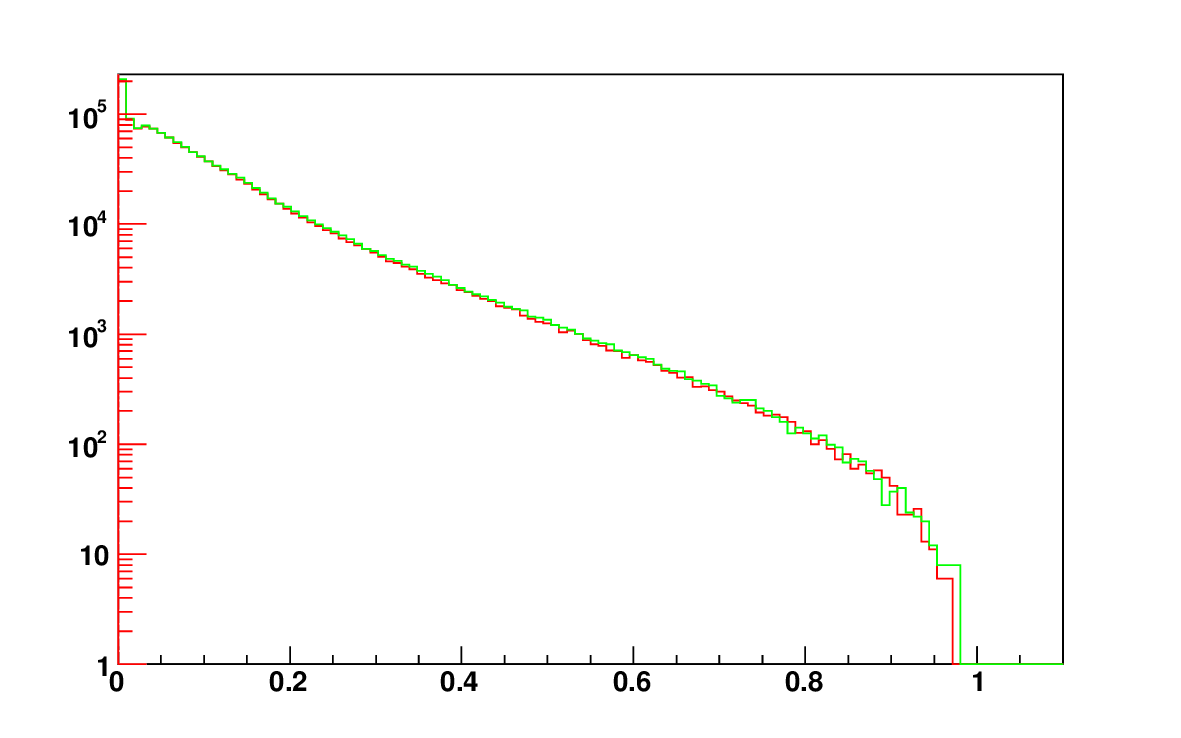}
}
\caption{ The spectrum of the $\gamma \gamma$ invariant mass, for the bremstrahlung photons in $Z \to \mu^+\mu^-$
decays. Events with two hard photons, both of energy above 1 GeV in the $Z$ rest 
frame are taken. Comparisons are shown for {\tt CEEX2} and {\tt CEEX1} (left plot), and {\tt CEEX2} and {\tt PHOTOS}
(right plot). The prediction from {\tt PHOTOS} is clearly superior for
the simulation of Higgs boson backgrounds. In the case of solutions based on YFS 
exponentiation, the second order matrix element must be taken into account. Fig.~\ref{plot:b} was obtained
from our example, {\tt examples/testing/Zmumu}, after adaptation of the center-of-mass energy 
(91.17 GeV) and test
energy threshold (1 GeV). Samples of 100 million events were used. 
See  {\tt examples/testing/README-plots} for technical details. Reference 
{\tt ROOT} files for {\tt KKMC CEEX} samples are, however, created outside of the {\tt PHOTOS} distribution package.
  \label{fig:gamgam}
}
\end{figure}

Agreement at this level is not seen in the differential distributions, see Fig.~\ref{fig:gamgam}. For example the spectrum of 
the two photon mass is quite different between the first and second order 
exponentiation result. This is of potential interest for background simulations 
for $H \to \gamma \gamma$. In contrast, the difference between the results from {\tt PHOTOS} and {\tt CEEX2} are much smaller. {\tt PHOTOS} exploits the first order matrix element 
in a better way than exponentiation. As a consequence it reproduces terms resulting in second order leading logarithms. This observation is important not only for 
the particular case of $Z$ decay but for the general case of double bremsstrahlung in any decay as well.

The numerical results collected here provide part of the program benchmarks. 
They are of limited but nonetheless  of  some physics interest as well.
 {\tt PHOTOS} provides only one step in the simulation chain: 
bremsstrahlung in decays of particles or resonances. One can ask 
the question of whether such a specialized unit is of interest and whether it is not better 
to provide the complete chain for ``truth physics simulation'' as a single simulation package. Obviously, this will
depend on particular needs. Final state QED corrections can be 
 separated from the remaining genuine electroweak corrections and 
in particular the initial state QED bremsstrahlung from quarks. 
One should bear in mind, that final state bremsstrahlung 
needs to be disentangled from detector acceptance dependencies and is usually not of 
interest in itself. This must be done e.g. to measure the properties of weak bosons.

One should bear in mind that even for QED FSR alone, the discussion of the physics 
precision of the simulation result requires further checks. In the case of 
$Z$ decays, Refs.~\cite{Golonka:2006tw,Golonka:2005pn} may not be enough.
With increasing precision, the estimation of uncertainty becomes dependent on 
the particular choice of details for the experimental cuts. Comparisons of different
calculations become important too. A good example of such work in the context of 
other measurements can be found in Refs.~\cite{Jadach:1995pd,Arbuzov:1996eq}; 
for the simulation of 
FSR QED corrections. The Monte Carlo programs collected for {\tt PHOTOS} generator 
tests are probably enough. 
A discussion of QED initial final state interference may follow the strategy presented
in \cite{Jadach:1999gz}. There, the question of experimental cuts must be included 
as well. Once these steps are finished, discussion 
of complete electroweak corrections in the context of realistic observables should 
be simplified.

Genuine weak corrections have to be taken into account separately.
Such solutions may be possible,  if together with the {\tt PHOTOS Interface},
 weak corrections are provided, for example, using the {\tt TAUOLA Interface}.
A discussion
of the complete electroweak corrections, as shown in Fig. 1a of  \cite{Adam:2008ge},
is not the purpose of our document. Let us point out however that electroweak 
non-QED corrections can be, in principle, installed into the {\tt PYTHIA 8.1} + {\tt PHOTOS} simulation 
using the e.g. {\tt TAUOLA} interface \cite{Davidson:2010rw}.
But for such a solution to be precise, further work is needed \cite{Bardin-private}.

Sizeable initial state QED corrections are usually embodied in  units 
simulating parton showers. This may need some experimental analysis as well. 
Experimental data from LEP1 were revisited by the DELPHI collaboration \cite{Abdallah:2010tk}.
Tension between data and the theoretical description was
mentioned. This may mean that the description of initial state QED bremsstrahlung 
at the LHC will need to be re-investigated using LHC data as well.
That is also why it  might be useful to keep initial state QED, final state QED and their interference corrections in separate modules. 


\section{Summary and outlook}
\label{sec:summary}
We have presented a new version  of {\tt PHOTOS} Monte Carlo. The part of 
{\tt PHOTOS} which operates on 
event records is now rewritten into {\tt C++} and an interface to the {\tt HepMC} event record 
is prepared. Interface to the {\tt HEPEVT} event record of {\tt FORTRAN} is provided as well.
The physics performance of the program is the
same, or better, than that of the {\tt FORTRAN/HEPEVT} version and better steering options are introduced. 
When an elementary decay is to be modified by {\tt PHOTOS}, 
it is first transformed to its rest frame. The $z$-axis is orientated along the decaying particle's mother's direction, 
as seen in this rest  frame. Such modification is 
necessary to calculate process dependent kernels
featuring the complete first order matrix element. 
The appropriate kernels explained in
Refs.~\cite{Golonka:2006tw,Nanava:2006vv,Nanava:2009vg} are installed
into the version, now fully in {\tt C}, of the internal algorithms of {\tt PHOTOS}. The necessary information is extracted
from the event record and  used.
No more  parts of the algorithm are left in {\tt FORTRAN}.
 The remaining {\tt FORTRAN} part of the code is for the optional interface to {\tt HEPEVT}.

{Finally let us point to ref.~\cite{Arbuzov:2012dx}. Thanks to this work, for LHC applications in $Z$ and $W$ decays,
the {\tt PHOTOS} Monte Carlo systematic error was established at  0.3\%, even   0.2\% for the case when matrix element corrections were activated. The estimation is valid
for complete final state radiative corrections, not for photonic bremsstrahlung
alone, even in the case when lepton pair emission is not taken into account.
}

\vskip 2 mm

\centerline{\large\bf Acknowledgements}

Useful discussions with P. Golonka during the early stage of project development and discussions 
with members of the ATLAS and CMS collaborations and the LCG/Genser team 
are acknowledged.
Nice atmosphere of Galileo Galilei Institute of Theoretical Physics  in Firenze where part of this 
work was performed is mentioned. 

Partial support by the Polish-French collaboration
no. 10-138 within IN2P3 through LAPP Annecy and 
during the years leading to  completion of this work is
also acknowledged.

\providecommand{\href}[2]{#2}


\newpage
\appendix

\section{Appendix: Interface to internal {\tt C} part of {\tt PHOTOS}}
\label{Interface to PHOTOS}

This appendix is addressed to developers of the interface,
and special users interested in advanced options of {\tt PHOTOS}.
The {\tt COMMON} blocks of {\tt FORTRAN} discussed below were used in the
program up  to version 3.53.
Starting from version 3.54 some of these {\tt COMMON}  blocks were preserved,
as the struct objects of {\tt C}. The following ones may be of particular interest for the
user:
{\tt PHOCOP, PHOKEY, PHOSTA}.
Names of variables and structs are not modified with respect to {\tt FORTRAN}, 
the  {\tt C++} definition style is however adopted: small letters are used 
for structs and their variable names.

The event record common block, {\tt HEPEVT}, is preserved but for use only in the {\tt FORTRAN} examples.  
The interface is available
through the classes {\tt PhotosHEPEVTEvent.h} and {\tt PhotosHEPEVTParticle.h}.
Let us recall once again, that internally in the {\tt C++} {\tt PHOTOS} code, the {\tt HEPEVT}-like  data type
is used in structs declaration.

\subsection{Common Blocks migrated to struct objects of C}

In the following let us list the original common blocks of {\tt FORTRAN} {\tt PHOTOS} which are in fact now replaced with struct objects of {\tt C}. 
To simplify 
and to preserve continuity of  the mathematical formulae shapes,
in many places  we  use  aliases to the struct elements  
which differ only in that they are  written in capital letters instead of small ones.
Note that relevant parameters  listed only here, can be set through
the appropriate accessors see Appendix \ref{subsection:other_methods}  for more explanations.

\begin{description}
\item[PHOCOP] coupling constant and related parameters.
    \begin{description}
	\item[ALPHA]  \textit{double} coupling constant $\alpha_{QED}$.
	\item[XPHCUT] \textit{double} minimal energy (in units of half of the decaying particle's mass) for photons to be explicitly generated.
    \end{description}
\end{description}

\begin{description}
\item[PHOKEY] keys and parameters controlling the algorithm options.
    \begin{description}
	\item[FSEC]   \textit{double} internal variable for algorithm options, the default is FSEC=1.0\; .
	\item[FINT]   \textit{double} maximum interference weight.
	\item[EXPEPS] \textit{double} technical parameter which blocks the crude level high photon multiplicity from configurations less probable than EXPEPS. The default is $10^{-4}$.
	\item[INTERF] \textit{bool} switch for interference, in the matrix element weight.
	\item[ISEC]   \textit{bool} switch for double bremsstrahlung generation.
	\item[ITRE]   \textit{bool} switch for bremsstrahlung generation up to a multiplicity of 4.
	\item[IEXP]   \textit{bool} switch for exponentiation mode.
	\item[IFTOP]  \textit{bool} switch for photon emission in top pair production in quark (gluon) pair annihilation.
	\item[IFW]    \textit{bool} switch for leading effects of the matrix element in leptonic $W$ decays.
    \end{description}
\end{description}

\begin{description}
\item[PHOSTA] Status information.
    \begin{description}
    \item[STATUS[ 10]]  \textit{int} Status codes for the last 10 errors/warnings
    that occurred.
    \end{description}
\end{description}

\begin{description}
\item[PHPICO] $\pi$ value definition.
    \begin{description}
    \item[PI]  \textit{double} $\pi$.
	\item[TWOPI]  \textit{double} $2*\pi$.
    \end{description}
\end{description}

\subsection{Routines}

In the following let us list routines which are called from the interface.

\begin{description}
\item[PHODMP] prints out the content of struct {\tt hep}. \\
  Return type: \textit{void} \\
  Parameters: none
\end{description}

\begin{description}
\item[PHOTOS\_MAKE\_C]  like {\tt PHOTOS\_MAKE} from the {\tt FORTRAN} part of the interface, but now in {\tt C}. \\
  Return type: \textit{void} \\
  Parameters:
  \begin{enumerate}
    \item \textit {int id} ID of the particle from which {\tt PHOTOS} starts 
processing. In the {\tt C++} case the importance of this parameter is limited 
as only one branch, reduced to the decay (process)  under consideration,  is in the {\tt hep} struct at a time. 
  \end{enumerate}
\end{description}

\begin{description}
\item[PHCORK] initializes kinematic corrections. \\
  Return type: \textit{void} \\
  Parameters:
  \begin{enumerate}
    \item \textit {int modcor} type of correction. See Ref.~\cite{Golonka:2005pn}  for details.
  \end{enumerate}
\end{description}

\begin{description}
\item[IPHQRK] enables/blocks (2/1) emission from quarks. \\
  Return type: \textit{int} \\
  Parameters: \textit{int}
\end{description}

\begin{description}
\item[IPHEKL] enables/blocks (2/1) emission in: $\pi^0 \rightarrow \gamma e^+ e^-$. \\
  Return type: \textit{int} \\
  Parameters: \textit{int}
\end{description}

\section{Appendix: User Guide}
\label{sec:User Guide}

\subsection{Installation}
\label{sec:Installation}
 
{\tt Photos C++ Interface} is distributed in the form of an archive containing source files and examples.
Currently only the Linux and Mac OS\footnote{For this case LCG configuration 
scripts explained in Appendix \ref{sec:autotools} have to be used.} operating systems are supported: other systems may be
supported in the future if sufficient interest is found.

The main interface library uses {\tt HepMC} \cite{Dobbs:2001ck} (version 2.03 or later) and requires that either
its location has been provided or compilation without {\tt HepMC} has been chosen as an option during the configuration step.
The later is only sufficient to compile the interface and to run the {\tt HEPEVT} example.

In order to run further examples located in the {\tt /examples} directory, {\tt HepMC} is required.
To run all of the available examples, it is  required to install:

\begin{itemize}
  \item {\tt ROOT} \cite{root-install-www} version 5.18 or later
  \item {\tt PYTHIA 8.2} \cite{Sjostrand:2007gs} or later\footnote{
        Examples can be adapted to use {\tt pythia8.1}. The necessary changes are explained in {\tt examples/README-PYTHIA-VERSIONS}. In fact, many of our numerical results stored in a code tar ball were obtained with older versions of 
{\tt PYTHIA}.}.
  \item {\tt MC-TESTER} \cite{Golonka:2002rz,Davidson:2008ma} version 1.24 or later.
        Do not forget to type {\tt make libHepMCEvent} after compilation of {\tt MC-TESTER} is done.
  \item {\tt TAUOLA} \cite{Davidson:2010rw} version 1.0.5 or later. {\tt TAUOLA} must be compiled with {\tt HepMC}.
\end{itemize}

In order to compile the {\tt PHOTOS} {\tt C++} Interface:
\begin{itemize}
 \item Execute {\tt ./configure} with the additional command line options:
   \subitem {\tt --with-hepmc=$<$path$>$} provides the path to the {\tt HepMC} installation directory. One can also set the {\tt HEPMCLOCATION} variable instead of using this directive. To compile the interface without {\tt HepMC} use {\tt --without-hepmc}
   \subitem {\tt --prefix=$<$path$>$} specifies the installation path. The {\tt include} and {\tt lib} directories will be copied there if {\tt make install} is executed later. If none has been provided, the default directory for installation is {\tt /usr/local}.
 \item Execute {\tt make}
 \item Optionally, execute {\tt make install} to copy files to the directory provided during configuration.
\end{itemize}

The {\tt PHOTOS} {\tt C++} interface will be compiled and the {\tt /lib} and {\tt /include} directories will contain the appropriate libraries and include files.

In order to compile the examples, compile the {\tt PHOTOS} {\tt C++} interface, enter the {\tt /examples} directory and:
\begin{itemize}
  \item Execute {\tt ./configure} to determine which examples can be compiled.
        Optional paths, required to compile additional examples and tests, can be provided as command line options
        (note that all of them are required for tests located in {\tt examples/testing} directory):
   \subitem {\tt --with-pythia8=$<$path$>$} provides the path to the {\tt Pythia8} installation
            directory. One can set the {\tt PYTHIALOCATION} variable instead of using this directive.
   \subitem {\tt --with-mc-tester=$<$path$>$} provides the path to the {\tt MC-TESTER} installation
            directory (the {\tt libHepMCEvent} must be compiled as well, see Ref.~\cite{Davidson:2008ma}
			for more details). One can set the {\tt MCTESTERLOCATION} variable instead of using this
			directive. This option
			implies that {\tt ROOT} has already been installed (since it is required by {\tt MC-TESTER}).
			The {\tt ROOT} directory {\tt bin} should be listed in the variable {\tt PATH} and the {\tt ROOT}
			libraries in {\tt LD\_LIBRARY\_PATH}.
   \subitem {\tt --with-tauola=$<$path$>$} provides the path to the {\tt TAUOLA} installation directory.
            One can set the {\tt TAUOLALOCATION} variable instead of using this directive.
  \item Execute {\tt make}
\end{itemize}

The {\tt /examples} directory will
contain the executable files for all examples that can be compiled and linked.
Their availability depends on the optional paths listed above.
If neither {\tt HepMC}, {\tt Pythia8}, {\tt Tauola} nor {\tt MC-TESTER} are accessible, 
only the {\tt HEPEVT} example will be provided.

\subsection{ LCG configuration scripts; available from version 3.1%
}
\label{sec:autotools}
For our project still another configuration/automake system was prepared
for use in LCG/Genser projects\footnote{We have used the expertise and advice
of Dmitri Konstantinov and Oleg Zenin in organization of configuration scripts
for our whole distribution tar-ball as well. Thanks to this choice, we hope, our solution
 will be compatible with ones in general use.} \cite{LCG,Kirsanov:2008zz}.

To activate this set of autotool based \cite{autotools} installation scripts,
enter the {\tt platform} directory and execute the {\tt use-LCG-config.sh} script.
Then, the installation procedure and the names of the configuration script parameters will differ from the ones 
described in our paper. Instruction given in the  './INSTALL' readme file, created by the {\tt use-LCG-config.sh} script,
should be followed. One can also execute {\tt ./configure --help}, which will 
list all options available for the configuration script.

Breif information on these scripts can be found in {\tt README} in the main directory as well.

\subsection{Elementary Tests}
\label{sect:elem}

The most basic test which should be performed, for our custom examples but also for a user's own generation chain, 
 is verification that the interface is installed correctly, 
photons are indeed added by the program and that energy momentum 
conservation is preserved\footnote{
We have performed such tests for {\tt HepMC} events obtained 
from {\tt PYTHIA 8.1}, {\tt PYTHIA 8.135}, {\tt PYTHIA 8.165}, {\tt PYTHIA 8.185} and {\tt PYTHIA 8.201}
using all configurations mentioned in this paper, all config files in {\tt examples} directory and
subdirectories of {\tt examples/testing}. Further  options for initializations 
(parton shower hadronization or QED bremsstrahlung on/off etc.) were also 
studied for different {\tt PYTHIA 8.1} versions. This was a necessary step 
in our program development. 

However, we do not document studies of 
{\tt Pythia} physics initialization  for all of its versions. 
That is why, distributions monitoring production processes obtained 
from distributed initialization for {\tt PYTHIA 8.201}, may differ from 
the  reference ones. See e.g. 
{\tt User Histograms}, plots {\tt mother-PT mother-eta},  in 
{\tt examples/testing/ScalNLO} or {\tt examples/testing/WmunuNLO}. }.

In principle, these tests have to be performed for any new hard 
process and after any new installation. This is to ensure that 
information is passed from the event record to the interface 
correctly and that physics information is filled into the {\tt HepMC} event 
in the expected manner. Misinterpretation of the event record content may result in 
faulty {\tt PHOTOS} operation.

\subsection{Executing Examples}

Once elementary tests are completed one can turn to the more advanced ones.
The purpose is not only to validate the installation but to demonstrate the
interface use.

The examples can be run by executing the appropriate {\tt .exe} file in the {\tt /examples} directory.
In order to run some more specific tests for the following processes:
$H \rightarrow \tau^+ \tau^-$, $ e^+ e^- \rightarrow t \bar t$,
$W \rightarrow e \nu_e$, $W \rightarrow \mu \nu_\mu$,
$Z \rightarrow e^+ e^-$, $Z \rightarrow \mu \mu$ or $Z \rightarrow \tau^+ \tau^-$,
$K_{0}^{S} \rightarrow \pi \pi$,
the main programs residing in the subdirectories of {\tt /examples/testing} should be executed.
Note that all paths listed as optional in Appendix \ref{sec:Installation} are required for these
tests to work.
In all cases the following actions have to be performed:

\begin{itemize}
  \item Compile the {\tt PHOTOS} {\tt C++} Interface. 
 \item  Check that the appropriate system variables are set. Execution of  the script \\
{\tt /configure.paths.sh} can usually perform this task; the configuration step 
announces this script.
  \item Enter the {\tt /examples/testing} directory. Execute {\tt make}. 
Modify {\tt test.inc} if needed.
  \item Enter the sub-directory for the particular process of interest
and execute {\tt make}.
\end{itemize}

The appropriate {\tt .root} files as well as {\tt .pdf} files generated by {\tt MC-TESTER}
will be created inside the chosen directory. One can execute 'make clobber' to
clean the directory. One can also execute 'make run' inside the {\tt /examples/testing}
directory to run all available tests one after another. Changes in source
code  can  be partly validated in this way.
Most of the tests are run using the executable {\tt examples/testing/photos\_test.exe}. The 
 $K_{0}^{S} \rightarrow \pi \pi$, $H \rightarrow \tau^+ \tau^-$ and $Z \rightarrow \tau^+ \tau^-$ examples 
require
{\tt examples/testing/photos\_tauola\_test.exe} to be run.
After generation, {\tt MC-TESTER} booklets will be produced,
 comparisons to the benchmark files will be shown.
A set of benchmark {\tt MC-TESTER} root files have been included with the interface
distribution. They are located in the subdirectories of {\tt examples/testing/}.
Note that for the $W \rightarrow e \nu_e$, 
$W \rightarrow \mu \nu_\mu$ and $Z \rightarrow \mu \mu$
examples,   differences higher than statistical error will show. 
This is because  photon symmetrization
was used in the benchmark files generated with {\tt KKMC}, and not in the ones 
generated with {\tt PHOTOS}.
In the case of {\tt KKMC} the generated photons are strictly ordered in energy. 
In the case of {\tt PHOTOS} they are not. Nonetheless, on average, 
the second photon has a smaller energy than the one written as the first
in the event record.

The comparison booklets can be useful 
to start new work or simply to 
validate new versions or new installations of the {\tt PHOTOS} interface.

In Appendix \ref{sec:User Configuration}, possible modifications to the  
example's settings are discussed. This may be interesting as an initial step for user's 
physics studies.  The numerical results of some of these tests are collected in Section \ref{sec:results}
and can be thus reproduced by the user.

\subsection{How to Run {\tt PHOTOS} with Other Generators}
If a user is building a large simulation system she or he may want to avoid
integration with our full configuration infrastructure and only load the libraries. 
For that purpose our stand-alone 
example {\bf examples/photos\_standalone\_example.exe} is a good starting point.

In order to link the libraries to the user's project, both the static libraries and shared objects are
constructed. To use the {\tt PHOTOS} interface in an external project, additional 
compilation directives are required. For the static libraries:
\begin{itemize}
  \item add {\tt -I<PhotosLocation>/include} at the compilation step,
  \item add {\tt <PhotosLocation>/lib/libPhotospp.a} as well as one or both of the libraries: 
            {\tt <PhotosLocation>/lib/libPhotosppHepMC.a} and
            {\tt <PhotosLocation>/lib/libPhotosppHEPEVT.a} to the linking step of your project.
\end{itemize}
For the shared objects:
\begin{itemize}
  \item add {\tt -I<PhotosLocation>/include} at the compilation step,
  \item add {\tt -L<PhotosLocation>/lib} along with {\tt -lPhotospp} as well as one or both of the libraries:
            {\tt -lPhotosppHepMC} and {\tt -lPhotosppHEPEVT} to the linking step.
  \item  {\tt PHOTOS} libraries must be provided for the executable; e.g. with the help of {\tt LD\_LIBRARY\_PATH}.
\end{itemize}
{\tt <PhotosLocation>} denotes the path to the {\tt PHOTOS} installation directory.
In most cases it should be enough to include within a users's program {\tt Photos.h} and {\tt PhotosHepMCEvent.h} (or any other header file for the class implementing abstract class {\tt PhotosEvent})
With that, the {\tt Photos} class can be used for configuration and {\tt PhotosHepMCEvent}
for event processing.

\subsubsection{Running {\tt PHOTOS C++ Interface} in a {\tt FORTRAN} environment}

For backward-compatibility with {\tt HEPEVT} event records, an interface has been prepared
allowing the {\tt PHOTOS C++ Interface} to be invoked from the {\tt FORTRAN} project. An example,
{\tt photos\_hepevt\_example.f}, has been prepared to demonstrate how {\tt PHOTOS} can be
initialized and executed from {\tt FORTRAN} code. Since {\tt PHOTOS} works in a {\tt C++} environment,  \\
{\tt photos\_hepevt\_example\_interface.cxx} must be introduced to invoke {\tt PHOTOS}.

Since version 3.54, {\tt PHOTOS} is fully in {\tt C++} and initialization can no longer
be performed from {\tt FORTRAN} code through the use of common blocks. 
In particular, information from the field 
{\tt QEDRAD} localized in {\tt  FORTRAN} times common block 
{\tt PHOQED } -- the extension of {\tt HEPEVT} is ignored. 
The  {\tt Photospp} initialization methods can be used easily instead.

Note that in the case of {\tt HEPEVT}, the {\tt PHOTOS} algorithm
has to modify the pointers (stored as integer variables) between mothers 
and daughters for all particles stored downstream of the added 
photons or lepton pairs. This part of the code was not replicated in full detail. 
Also, most of
our tests were performed only on cases where the modified decay was the last one in the event record, thus  shifting consecutive entries was not necessary.
We do not foresee the use of the program and its development for circumstances distinct from these.

\section{Appendix: User Configuration}
\label{sec:User Configuration}

\subsection{Suppress Bremsstrahlung}
\label{section:suppress}

In general, {\tt PHOTOS} will attempt to generate bremsstrahlung for every 
branching point in the event record. This is of course not always appropriate.
Already inside the internal {\tt C} part of {\tt PHOTOS}, 
bremsstrahlung is normally prevented for vertices involving gluons or quarks 
(with the exception of top quarks).

This alone is insufficient. By default we suppress bremsstrahlung
generation for vertices like $l^\pm \to l^\pm \gamma$ because a
``self-decay'' is unphysical. We cannot request that all incoming
and/or outgoing lines are on mass shell, because it is not the case in
cascade decays featuring intermediate states of sizeable width. If a
parton shower features a vertex with $l^\pm \to l^\pm \gamma$ with the
virtuality of the incoming $l^\pm$ matching the invariant mass of the
outgoing pair then the action of {\tt PHOTOS} at this vertex will
introduce an error.  This is prevented by forbidding bremsstrahlung
generation at vertices where one of the decay products has a flavor
which matches the flavor of an incoming particle.

Some exceptions to the default behavior may be necessary. For example
in cascade decays, the vertex $\rho \to \rho \pi$ may require the
{\tt PHOTOS} algorithm to be activated.

Methods to re-enable these previously prevented cases or to prevent generation in special
branches have been introduced and are presented below. \\ \\

\begin{itemize}
 \item {\tt Photos::suppressBremForDecay(daughterCount, motherID, d1ID, d2ID, ...)} \hfill \\
       The basic method of channel suppression. The number of daughters,
	   {\tt PDGID} of the mother and the list of {\tt PDGID}s of daughters must be provided.
	   There is no upper limit to the number of daughters.
	   If a decay with the matching pattern is found, {\tt PHOTOS} will skip the decay.
           The decay will be skipped if it contains additional photons or other particles,
           as long as all of the particles from the pattern are present in the list of daughters.
 \item {\tt Photos::suppressBremForDecay(0, motherID)} \hfill \\
       When only the {\tt PDGID} of the mother is provided, (the {\it daughterCount} is 0) {\tt PHOTOS} will skip all decay channels
	   of this particle. 
 \item {\tt Photos::suppressBremForBranch(daughterCount, motherID, d1ID, d2ID, ...)} \hfill \\
       {\tt Photos::suppressBremForBranch(0, motherID)} \hfill \\
       The usage of this function is similar to the two cases of the previous function. The difference is
	   that {\tt PHOTOS} will skip not only the corresponding channel,
	   but also all consecutive decays of its daughters, making {\tt PHOTOS} skip the entire branch
	   of decays instead of just one.
 \item {\tt Photos::suppressAll() }
       All branchings will be suppressed except those that are forced using the methods
	   described in the next section.
 \item \textbf{Example:} \hfill \\
{\tt Photos::suppressBremForDecay(3, 15, 16, 11, -12); } \\
{\tt Photos::suppressBremForDecay(2, -15, -16, 211); } \\
{\tt Photos::suppressBremForDecay(0, 111); } \\
\emph{If the decays $\tau^- \rightarrow \nu_\tau e^- \bar \nu_e$ or
      $\tau^+ \rightarrow \bar \nu_\tau \pi^+$ are found, they will be skipped by {\tt PHOTOS}{%
      \footnote{Note that the first line of this example states that any decays of $\tau^-$
      that contain $\nu_\tau$, $e^-$ and $\bar \nu_e$ will be skipped regardless of how many other particles are in the decay.
      So, for example, $\tau^- \rightarrow \nu_\tau e^- \bar \nu_e \gamma$ will be skipped as well. It is important to realize that excluding 
$\tau^-  \rightarrow \nu_\tau \pi^-$ leads to exclusion of
$\tau^- \rightarrow \nu_\tau \pi^-\pi^0$ as well. If it is not required it must be allowed separately.}}.
	  In addition, all decays of $\pi^0$ will also be skipped. Note, that the minimum
	  number of parameters that must be provided is two - the number of daughters
	  (which should be zero if suppression for all decay channels of the particle is chosen) 
	  and the mother {\tt PDGID}.} \\ \\
{\tt Photos::suppressBremForBranch(2, 15, 16, -213); } \\
\emph{When the decay $\tau^- \rightarrow \nu_\tau \rho^-$ is found, it will be skipped by
      {\tt PHOTOS} along with the decays of   $\rho^-$ 
(in principle also $\nu_\tau$) and all
	  their daughters. In the end, the whole decay tree starting with
	  $\tau^- \rightarrow \nu_\tau \rho^-$ will be skipped.}
\end{itemize}

In future, an option to suppress a combination of consecutive branches may be introduced.
For example if bremsstrahlung in leptonic $\tau$ decays is generated by
programs prior to {\tt PHOTOS}, and the decay is stored in {\tt HepMC} as the cascade
$\tau^\pm \to W^\pm \nu$, $W^\pm \to l^\pm \nu$, {\tt PHOTOS} must
be prevented from acting on both vertices, but only in cases when they are present one after another.
One can also think of another {\tt PHOTOS} extension. 
If a vertex $q \bar q \to l^\pm l^\mp$ is found, then it should not be ignored 
that intermediate state can be then attributed, 
$q \bar q \to Z/\gamma^* \to l^\pm l^\mp$,
and used for matrix element calculation.

\subsection{Force {\tt PHOTOS} Processing }
\label{section:force}

Forcing {\tt PHOTOS} to process a branch can be used in combination with
the suppression of all branches i.e. to allow selection of only a particular
processes for bremsstrahlung generation.

Forced processing using the methods below has higher priority than the suppression described
in the previous section, therefore even if both forcing and suppressing of the same
branch or decay is done (regardless of order), the processing will not be
suppressed.

\begin{itemize}

 \item {\tt Photos::forceBremForDecay(daughterCount, motherID, d1ID, d2ID, ...)} \hfill \\
       {\tt Photos::forceBremForDecay(0, motherID)} \hfill \\
       The usage of this function is similar to {\tt Photos::suppressBremForDecay(...)}
	   described in the previous section. If a decay with the matching pattern is found,
	   {\tt PHOTOS} will be forced to process the corresponding decay, even if it was suppressed
	   by any of the methods mentioned in the previous section.
 \item {\tt Photos::forceBremForBranch(daughterCount, motherID, d1ID, d2ID, ...)} \hfill \\
       {\tt Photos::forceBremForBranch(0, motherID)} \hfill \\
       The usage is similar to the above functions. The difference is
	   that {\tt PHOTOS} will force not only the corresponding channel,
	   but also all consecutive decays of its daughters, making {\tt PHOTOS} process the entire branch
	   of decays instead of just one. This method can activate part of the later branch previously prevented.
 \item \textbf{Example:} \hfill \\
{\tt Photos::suppressAll(); } \\
{\tt Photos::forceBremForDecay(4, 15, 16, -211, -211, 211); } \\
{\tt Photos::forceBremForDecay(2, -15, -16, 211); } \\
{\tt Photos::forceBremForBranch(0, 111); } \\
\emph{Since suppression of all processes is used, only the listed decays will be processed,
      these are $\tau^- \rightarrow \nu_\tau \pi^- \pi^- \pi^+$, $\tau^+ \rightarrow \bar \nu_\tau \pi^+$
      and all instances of the decay of $\pi^0$ and its descendants.}
\end{itemize}

\subsection{Use of the {\tt processParticle} and {\tt processBranch} Methods}
\label{PHOTOSgun}

In Section~\ref{sect:Outline} the algorithm for processing a whole
event record is explained and is provided through the {\tt process()}
method.  To process a single branch in the event record, in a way
independent of the entire event, a separate method is provided.

\begin{itemize}
  \item {\tt Photos::processParticle(PhotosParticle *p) } \hfill \\
		The main method for processing a single particle decay vertex. A pointer to a particle must
		be provided. Pointers to mothers and daughters of this particle should be
		accessible through this particle or its event record.
		From this particle a branch containing its mothers and daughters
		will be created and processed by {\tt PHOTOS}.
  \item {\tt Photos::processBranch(PhotosParticle *p) } \hfill \\
		Usage is similar to the above function. When a pointer to a particle is provided,
		{\tt PHOTOS} will process the whole decay branch starting from the particle provided.
\end{itemize}

An example, {\tt single\_photos\_gun\_example.c}, is provided in the directory {\tt /examples}
showing how this functionality can be used to process the decay of selected particles.
$Z^0 \rightarrow \tau^+ \tau^-$ decays are generated and the event record is traversed
searching for the first $\tau^-$ particle in the event record.
Instead of processing the whole event, only the decay of a $\tau^-$ is processed by {\tt PHOTOS}.

\subsection{Lepton pair emission and event record momentum unit}
\label{sec:units}

For the purpose of pair emission, introduced for the first time   with {\tt Photos} version
3.57, the information about the momentum unit
(either {\tt GEV} or {\tt MEV}) used by the event has to be provided.
For other {\tt Photos} applications this information is not needed.

In case of the {\tt HepMC} event record, this information is automatically obtained
from the event. For {\tt HEPEVT} events, {\tt GEV} is assumed. For all other
interfaces, the unit is undefined and has to be set in the event record
interface by calling \\ {\tt Photos::setMomentumUnit(MomentumUnits unit); }\\
(e.g. {\tt Photos::setMomentumUnit(Photos::MEV); }).
\\ See
constructors for the {\tt PhotosHepMCEvent} or {\tt PhotosHEPEVEvent} class for
further examples.

To select available emission from {\tt PHOTOS} use:

\begin{itemize}
  \item {\tt Photos::setPairEmission(bool flag); } \\
        Turn on or off emission of pairs (electrons or muons). Default is off.
  \item {\tt Photos::setPhotonEmission(bool flag); } \\
        Turn on or off emission of photons. Default is on.
\end{itemize}

Tests and implementation of final state radiation 
pair emission, presently follow formulae 1 and 11 from Ref.~\cite{Jadach:1993wk}. 
The agreement, when pair emission phase space was restricted to the soft region, was  at the   2-5 \% 
level of the pair effect (which itself is at the 0.1 \% level of the cross section). 
The phase space  is parametrized without any mass or other approximations. This feature 
was checked separately with special runs (matrix element removed) of 100 Mevt samples.

The matrix element used for pair emission in decays is easy to improve. 
Dependence of four-momenta of final state particles is coded 
explicitly. Further work \cite{Sadykov} will be devoted to this task.
\subsection{Logging and Debugging}
\label{App:Logging}
This section describes the basic functionality of the logging and debugging tool.
For details on its content we address the reader to comments in the {\tt /src/utilities/Log.h} header file.

Let us present however a general scheme of the tool's
functionality.  The {\tt PHOTOS} interface allows control over the
amount of message data displayed during program execution and
provides a basic tool for memory leak tracking. The following initialization
functions can be used in a user's main program.  
The  {\tt Log.h} header has to be then incuded.
\begin{itemize}
  \item {\tt Log::LogPhlupa(int from, int to) } \\
        Turns logging of debug messages from the {\tt C} part of the program on and off.
        Parameters of this routine specify the range of debug codes for the {\tt phlupa} routine.
  \item {\tt Log::Summary() } - Displays a summary of all messages from {\tt C++} part of the code.
  \item {\tt Log::SummaryAtExit() } - Displays the summary at the end of a program run.
  \item {\tt Log::LogInfo(bool flag) } \\
        {\tt Log::LogWarning(bool flag) } \\
        {\tt Log::LogError(bool flag) } \\
        {\tt Log::LogDebug(int s, int e) } \\
        {\tt Log::LogAll(bool flag)} \\
        Turns the logging of \textit{info}, \textit{warning}, \textit{error} and \textit{debug} messages on or off depending
        on the flag value being true or false respectively. In the case of \textit{debug} messages - the range of message codes
        to be displayed must be provided. By default, only \textit{debug} messages
        (from 0 to 65535) are turned off. If the range is negative ($s>e$) \textit{debug} messages
        won't be displayed. The last method turns displaying all of the above messages on and off.
\end{itemize}

With  {\tt Log::LogDebug(s,e)}  messages of $s$ to $e$ range, 
will be printed at execution time, in particular:

\begin{itemize}
  \item Debug(0)    - seed used by the random number generator
  \item Debug(1)    - which type of branching was found in {\tt HepMC}
                     (regular or a case without an intermediate particle, for details see {\tt PhotosBranch.cxx})
  \item Debug(700)  - execution of the branching filter has started
  \item Debug(701)  - branching is forced
  \item Debug(702)  - branching is suppressed
  \item Debug(703)  - branching is processed (i.e. passed to the filter)
  \item Debug(900)  - started check of Matrix Element (ME) calculation for the  channel 
  \item Debug(901)  - ME channel value obtained
  \item Debug(902)  - final ME channel value after checking all flags
  \item Debug(2)    - execution of the branching filter was completed
  \item Debug(1000) - the number of particles sent to and retrieved from internal {\tt PHOTOS} event record.
\end{itemize}
 
The option {\tt Log::SetWarningLimit(int limit)} results in 
only the first {\tt `limit'} warnings being displayed. The default for {\tt limit} is 100. 
If {\tt limit}=0 is set, then there are no limits on the number of warnings to be displayed.

The memory leak tracking function allows checking of whether all memory allocated within {\tt PHOTOS Interface}
 is properly released. However, using the debug option significantly increases the amount of time needed for 
each run. Its  use is therefore recommended  for debugging purposes only. In order to use this option
 modify {\tt make.inc} in the main directory by adding the line: \\ 
 {\tt DEBUG = -D"\_LOG\_DEBUG\_MODE\_" } \\ 
Recompile the interface.
Now, whenever the program is executed a table will be printed at the end of the run,
listing all the pointers that were not freed, along with the memory they consumed.
If the interface works correctly without any memory leaks, one should get an empty table.

It is possible to utilize this tool within a user's program; however there are a few limitations.
The debugging macro from "Log.h" can create compilation errors if one compiles
it along with software which has its own memory management system (e.g. {\tt ROOT}).
To make the macro work within a user's program, ensure that {\tt Log.h} is the last header file
included in the main program.
It is enough to  compile the program with the {\tt -D"\_LOG\_DEBUG\_MODE\_"} directive added,
or {\tt \#define \_LOG\_DEBUG\_MODE\_} placed within the program before inclusion of
 the {\tt Log.h} file%
\footnote{Note that {\tt Log.h} does not need to be included within
the user's program  for the memory leak tracking tool to be used only for the {\tt PHOTOS} interface.
}.

\subsection{Other User Configuration Methods}
\label{subsection:other_methods}

The following auxiliary methods are prepared. They are useful for initialization 
or are introduced for backward compatibility.

\begin{itemize}
  \item {\tt Photos::setRandomGenerator(double (*gen)()) } {\it  installed in {\tt PHOTOS 3.52}}\\
        Replace random number generator used by {\tt Photos}.
        The user provided generator   must return a {\tt double} between 0 and 1. 
        {\tt Photos::setRandomGenerator(NULL)} will reset the program back to  
        the default generator, which is a copy of {\tt RANMAR}~\cite{James:1988vf,marsaglia:1987}.
  \item {\tt Photos::setSeed(int iseed1, int iseed2)} \\
        Set the  seed values for our copy of the random number generator {\tt RANMAR} \cite{James:1988vf,marsaglia:1987}.
  \item {\tt Photos::maxWtInterference(double interference)} \\
        Set the maximum interference weight. The default, 2, is adopted to decays where at most two charged decay products
        are present\footnote{For 
        the decays like $J/\psi \to 5\pi^+ 5\pi^-$ a higher value, at least equal to the number of charged decay 
        products, should be set. The algorithm performance will slow down linearly with the  maximum interference weight but all 
        simulation results will remain unchanged.   
        } and no matrix element based kernel is used\footnote{Also in this case a higher  than default 2 should be used.}.
  \item {\tt Photos::setInfraredCutOff(double cut\_off)} \\
        Set the minimal energy (in units of decaying particle mass)
        for photons to be explicitly generated.
  \item {\tt Photos::setAlphaQED(double alpha)} \\
        Set the coupling constant, alpha QED.
  \item {\tt Photos::setInterference(bool interference)} \\
        A switch for interference, matrix element weight.
  \item {\tt Photos::setDoubleBrem(bool doub)} \\
        Set double bremsstrahlung generation.
  \item {\tt Photos::setQuatroBrem(bool quatroBrem)} \\
        Set bremsstrahlung generation up to a multiplicity of 4.
  \item {\tt Photos::setExponentiation(bool expo)} \\
        Set the exponentiation mode.
  \item {\tt Photos::setCorrectionWtForW(bool corr)} \\
         A switch for leading effects of the matrix element (in leptonic $W$ decays)
  \item {\tt Photos::setMeCorrectionWtForScalar(bool corr)} \\
         A switch for complete effects of the matrix element (in scalar to two scalar decays) {\it  installed in {\tt PHOTOS 3.3}.}
  \item {\tt Photos::setMeCorrectionWtForW(bool corr)} \\
         A switch for complete effects of the matrix element (in leptonic decays of $W$'s produced from anihilation of light fermions) {\it  installed in {\tt PHOTOS 3.2} }
  \item {\tt Photos::setMeCorrectionWtForZ(bool corr)} \\
         A switch for complete effects of the matrix element (in leptonic $Z$ decays) {\it  installed in {\tt PHOTOS 3.1} }
 \item {\tt Photos::setTopProcessRadiation(bool top)} \\
Set photon emission in top pair production in quark (gluon) pair annihilation
and in top decay.
  \item {\tt Photos::initializeKinematicCorrections(int flag)} \\
        Initialize kinematic corrections necessary to avoid consequences of rounding errors.
  \item {\tt Photos::forceMassFrom4Vector(bool flag)}  \\
        By default, for all particles used by {\tt PHOTOS}, 
        mass is re-calculated and $\sqrt{E^2-p^2}$ is used. 
        If {\tt flag=false}, the particle mass stored in the  event record 
        is used. The choice may be important for the control 
        of numerical stability in the case of very light stable particles, but may be incorrect for decay products 
        themselves of non-negligible width.
  \item {\tt Photos::forceMass(int pdgid, double mass)} {\it  installed in {\tt PHOTOS 3.4}} \\
        For particles of {\tt PDGID} (or {\tt -PDGID})  to be processed by {\tt PHOTOS},
        the mass value attributed  by the user will be used instead of the one calculated
        from the 4-vector. Note that if both {\tt forceMass} and {\tt forceMassFromEventRecord} is
        used for the same {\tt PDGID}, the last executed function will take effect.
        Up to version 3.51, the option is active if {\tt forceMassFrom4Vector = true} (default).
        From version 3.52, the option works regardless of the setting of {\tt forceMassFrom4Vector}.
  \item {\tt Photos::forceMassFromEventRecord(int pdgid)} {\it  installed in {\tt PHOTOS 3.4}} \\
        For particles of {\tt PDGID} (or {\tt -PDGID}) to be  processed by {\tt PHOTOS},
        the mass value taken from the event record will be used instead of the one
        calculated from the 4-vector. Note that if both {\tt forceMass} and {\tt forceMassFromEventRecord} is
        used for the same {\tt PDGID}, the last executed function will take effect.
        Up to version 3.51, the option is active if {\tt forceMassFrom4Vector = true} (default).
        From version 3.52, the option works regardless of the setting of {\tt forceMassFrom4Vector}.
  \item {\tt Photos::createHistoryEntries(bool flag, int status)} {\it  installed in {\tt PHOTOS 3.4}} \\
        If set to {\tt true}, and if the event record format allows,
        {\tt Photos} will store history entries consisting of particles
        before processing\footnote{In case of {\tt HepMC}, it creates copies
        of all particles on the list of outgoing particles in vertices where
        the photon was added and will be added at the end of the list.}.
        History entries will have status codes equal to {\tt status}.
        The value of {\tt status} will also be added to the list of status
        codes ignored by {\tt Photos} (see {\tt Photos::ignoreParticlesOfStatus}).
        An example is provided in  {\tt photos\_pythia\_example.cxx}.
  \item {\tt Photos::ignoreParticlesOfStatus(int status)} Decay products with the status code
        {\tt status} will be ignored when checking momentum conservation and will not be passed
        to the algorithm for generating bremsstrahlung.
  \item {\tt Photos::deIgnoreParticlesOfStatus(int status)} Removes {\tt status} from
        the list of status codes created with {\tt Photos::ignoreParticlesOfStatus}.
  \item {\tt bool Photos::isStatusCodeIgnored(int status)} Returns {\tt true} if {\tt status}
        is on the list of ignored status codes.
  \item {\tt Photos::setMomentumConservationThreshold(double momentum\_{}conservation\_{}threshold)} \\
        Threshold relative to the difference of the sum of the 4-momenta of incoming and outgoing
        particles. The default value is 0.1. If larger energy-momentum non-conservation
        is found then photon generation is skipped in the vertex\footnote{
        In the past, momentum conservation was checked using the standard method of {\tt HepMC}.
        That effectively meant that {\it only} momentum, but not energy was checked.
        This turned out to be insufficient in some rare cases.}.
  \item {\tt Photos::iniInfo()} \\
        The printout performed with  {\tt Photos::initialize()}  will exhibit 
        outdated information
        once the methods listed above are applied. The reinitialized data can be printed
        using the {\tt Photos::iniInfo()} method.
        The same format as {\tt Photos::initialize()}  will be used.		
\end{itemize}

\subsection{Creating Advanced Plots and Custom Analysis}
\label{App:Plots}

In Section \ref{sec:results}, we have presented results of a non-standard
analysis performed by {\tt MC-TESTER}. Figure \ref{fig:lineshape} has been
obtained using a custom {\tt UserTreeAnalysis} located in the {\tt ZeeAnalysis.C} file
residing in the {\tt examples/testing/Zee} directory. This file serves as an
example of how custom analysis can be performed and how new plots can be
added to the project with the help of {\tt MC-TESTER}.

The basic {\tt MC-TESTER} analysis contains methods used by pre-set examples
in the subdirectories of {\tt examples/testing} to focus on at most one or two 
sufficiently hard photons from all the photons generated
by {\tt PHOTOS}. Its description and usage have already been documented in \cite{Davidson:2008ma}.
The content of {\tt ZeeAnalysis.C} is identical to the default {\tt UserTreeAnalysis}
of {\tt MC-TESTER} with the only addition being a method to create
the previously mentioned plot.

In order to create the $t \bar t$ example, an additional routine had to be added to {\tt photos\_test.c}.
Since {\tt MC-TESTER} is not designed to analyze processes involving
multiple incoming particles, we have used a method similar to that previously
used in the {\tt FORTRAN} examples, {\tt LC\_Analysis}, mentioned in \cite{Golonka:2002rz}, Section 6.1.
This routine, {\tt fixForMctester}, transforms the $X Y \rightarrow t \bar t$
process to the $100 \rightarrow t \bar t$ process,
where the momentum of the
special particle $100$ is $X + Y$. With this modification, {\tt MC-TESTER} can be set
up to analyze the particle $100$ in order to get a desirable result.

For more details regarding the plots created for this documentation, see
{\tt README-plots} located in the {\tt examples/testing/} directory.


\end{document}